\documentclass[preprint,11pt,authoryear]{elsarticle}
\usepackage{setspace}
\usepackage{natbib}
\usepackage{enumitem}
\usepackage{amsthm}
\usepackage{array}

\usepackage{booktabs}
\usepackage{multirow}
\usepackage[T1]{fontenc}
\usepackage{amssymb}
\usepackage{appendix}
\usepackage{amsmath}
\usepackage{longtable}
\usepackage[utf8]{inputenc}
\usepackage[T1]{fontenc}
\usepackage{xcolor}
\usepackage{amsmath,amssymb,amsfonts}
\usepackage{textcomp}
\usepackage{float}
\usepackage{array}
\usepackage{booktabs} 
\usepackage{bm}
\usepackage{epsfig}
\usepackage{soul}
\usepackage{pgf}
\usepackage{etoolbox} 
\usepackage{placeins}
\usepackage{multirow}
\usepackage{bigstrut}
\usepackage{comment}
\usepackage{rotating}
\usepackage{pdflscape}
\usepackage{amsmath} 
\usepackage{lscape}
\usepackage{subcaption}
\usepackage{setspace}
\usepackage{amssymb}
\usepackage{amsmath}
\usepackage{relsize}
\usepackage{url}
\usepackage{algorithm}
\usepackage{algpseudocode}
\usepackage{natbib}
\usepackage{graphicx}
\usepackage{multirow}
\usepackage{multicol}
\usepackage{comment}
\usepackage{booktabs} 
\usepackage{array} 
\usepackage{balance} 
\usepackage{soul}
\usepackage{siunitx}
\usepackage{graphicx}
\usepackage{float}
\usepackage{subcaption} 
\usepackage{float}
\usepackage{xcolor}
\usepackage{comment}
\usepackage{graphicx}
\usepackage{xcolor}
\usepackage{tikz}
\usetikzlibrary{positioning, shapes.geometric, arrows.meta, calc}

\usetikzlibrary{positioning,fit}

\usepackage{pifont}

\newcommand{\cmark}{\textcolor{green!60!black}{\ding{51}}}
\newcommand{\xmark}{\textcolor{red}{\ding{55}}}

\newcolumntype{C}[1]{>{\small\centering\arraybackslash}m{#1}}
\newcolumntype{L}[1]{>{\small\raggedright\arraybackslash}m{#1}}

\usepackage{array}

\usepackage[dvipsnames]{xcolor}
\usepackage{pifont}
\usepackage{xurl}
\usepackage[margin=2.5cm]{geometry}
\allowdisplaybreaks
\usepackage[final]{hyperref}

\usepackage{titlesec}

\titleclass{\subsubsubsection}{straight}[\subsubsection]

\newcounter{subsubsubsection}[subsubsection]
\renewcommand\thesubsubsubsection{\thesubsubsection.\arabic{subsubsubsection}}

\titleformat{\subsubsubsection}
  {\normalfont\normalsize\bfseries}
  {\thesubsubsubsection}{1em}{}

\titlespacing*{\subsubsubsection}
  {0pt}{3.25ex plus 1ex minus .2ex}{1.5ex plus .2ex}
\titleformat{\paragraph}[runin]
  {\normalfont\normalsize\bfseries}
  {\theparagraph}{1em}{}
\titlespacing*{\paragraph}
  {0pt}{3.25ex plus 1ex minus .2ex}{1em}

\makeatletter
\providecommand*{\toclevel@subsubsubsection}{4}
\makeatother

\makeatletter
\providecommand*{\toclevel@subsubsubsection}{4}
\makeatother

\journal{a journal}

\definecolor{myblue}{HTML}{000B4F}
\definecolor{myblue2}{HTML}{20368F}
\definecolor{myblue3}{HTML}{829CD0}
\definecolor{myblue4}{HTML}{C9D6F0} 
\definecolor{myblue5}{HTML}{E6ECFA} 
\definecolor{myborder}{HTML}{1C2A5A}

\begin{document}

\begin{frontmatter}

\title{School network reorganization under educational and spatial constraints using classical and quantum optimization}

\author[aff1]{Alessia Ciacco}
\ead{alessia.ciacco@unical.it}

\author[aff2]{Luigi Di Puglia Pugliese}
\ead{luigi.dipugliapugliese@icar.cnr.it}

\author[aff1]{Francesca Guerriero}
\ead{francesca.guerriero@unical.it}

\affiliation[aff1]{organization={Department of Mechanical, Energy and Management Engineering, University of Calabria},
           city={Rende},
            postcode={87036},
            state={CS},
            country={Italy}}
            
\affiliation[aff2]{organization={Istituto di Calcolo e Reti ad Alte Prestazioni, Consiglio Nazionale delle Ricerche (ICAR-CNR)},
           city={Rende},
           postcode={87036},
           state={CS},
           country={Italy}}

\begin{abstract}
School network reorganization is a strategic planning problem that requires balancing demographic trends, territorial accessibility, educational requirements, and institutional constraints while ensuring an efficient allocation of public resources. This paper proposes an optimization framework for school dimensioning decisions based on a novel Integer Linear Programming formulation integrating geographical, administrative, and educational criteria. A synthetic benchmark generator is introduced to evaluate the scalability and computational performance of the model on artificial instances, while a real-world case study involving the complete public school network of the Calabria region (Italy) is conducted using actual institutional, territorial, and demographic data. 
Furthermore, the model is implemented within a hybrid quantum optimization environment. 
The results show that the proposed formulation can be effectively solved by exact classical optimization and can also be represented and evaluated within a hybrid quantum-classical optimization framework. The computational experiments provide a proof-of-concept assessment of the applicability of hybrid quantum optimization to the considered school aggregation problem, while also highlighting the current computational advantage of classical optimization and the limitations of the tested benchmark instances. 

\end{abstract}

\begin{keyword}
School network reorganization, Hybrid quantum optimization, Facility Location Problem, Real-world case study, Territorial sustainability.
\end{keyword}

\end{frontmatter}
\section{Introduction}
School network reorganization has become a major policy challenge in many countries worldwide. Educational systems are increasingly required to adapt their organizational structures to profound demographic, economic, and territorial transformations. Declining student populations, fiscal constraints, and increasing pressure to improve the efficiency of public expenditure have led governments to redesign school networks through mergers, consolidations, and the reallocation of administrative responsibilities. These interventions, commonly referred to as school network reorganization or school dimensioning, aim to optimize the allocation of educational resources while maintaining adequate service provision across territories.
However, school network reorganization is far from being a purely administrative or financial exercise. Schools represent essential public infrastructures whose role extends beyond education itself. They contribute to social cohesion, local development, and equal opportunities, particularly in rural and geographically isolated communities. Consequently, decisions aimed at reducing the number of autonomous school institutions inevitably involve multiple and often conflicting objectives. 
Designing an efficient school network therefore requires balancing economic sustainability with educational accessibility and territorial equity.
These challenges have become particularly relevant in Europe, where demographic decline is reshaping the geography of public services. During the last decades, persistently low fertility rates, population ageing, and internal migration toward metropolitan areas are reducing the number of school-age children, particularly in rural and peripheral regions. In 2024, the total fertility rate in the European Union fell to 1.34 children per woman, well below the replacement level of 2.1, while Eurostat projects that the EU population will decline by approximately 53 million inhabitants ($-11.7\%$) by the end of the century, accompanied by a continuous reduction in the share of children and young people.\footnote{Source: \citet{Eurostat2026}.} 
Within this European context, Italy represents one of the European countries most severely affected by demographic decline. In 2024, the country's total fertility rate was only 1.18 children per woman, well below the European Union average of 1.34, making Italy one of the countries with the lowest fertility levels in Europe. The reduction in fertility has been accompanied by a persistent decline in the young population. The share of residents younger than 15 years decreased from 14.2\% in 2004 to 11.9\% in 2025, compared with a decline from 16.2\% to 14.4\% in the European Union. Consequently, Italy now exhibits one of the lowest proportions of children in Europe, with demographic decline being particularly severe in the southern regions. \footnote{Source: \citet{ECItaly2025}.} This demographic transition has generated increasing pressure on educational systems, forcing public authorities to reconsider the spatial organization of schools while preserving equal access to education. 
Recent legislative reforms have progressively strengthened school dimensioning policies by encouraging the consolidation of autonomous institutions in order to improve administrative efficiency and optimize public expenditure. 
Although demographic decline affects the entire country, its effects are considerably more pronounced in Southern Italy, where long-term population loss and ageing have become major structural challenges for the provision of essential public services. Among the southern regions, Calabria represents one of the territories most severely affected by demographic decline, population ageing, and outward migration. The region is characterized by a large number of inner and peripheral municipalities experiencing a continuous reduction in the school-age population, making the provision of educational services particularly challenging.\footnote{Source: \citet{ISTAT2025Demography}.} These territorial characteristics considerably reduce accessibility to essential public services and increase the strategic importance of local schools as institutions supporting social cohesion and territorial resilience. In many municipalities, schools represent not only educational facilities but also key components of the local institutional infrastructure. Consequently, decisions concerning school aggregation may produce effects extending well beyond administrative efficiency, influencing accessibility, local development, and the long-term sustainability of fragile communities. 

School network reorganization can be formulated as a complex combinatorial optimization problem involving multiple and potentially conflicting objectives. The structure of the problem is strongly influenced by the institutional and regulatory characteristics of the educational system under consideration. 

This paper proposes an Integer Linear Programming (ILP) model for regional school network reorganization in the Italian context that jointly incorporates institutional compatibility, administrative regulations, school aggregation decisions, road-network travel-time accessibility, territorial fragility, and school capacity constraints. 

The model addresses the school aggregation component of school network reorganization, namely the selection of institutions that retain administrative autonomy and the assignment of eligible institutions to surviving school hubs. It does not constitute a complete end-to-end school network reorganization model including all possible planning decisions, such as school opening or closure, facility relocation, capacity expansion, transport planning, multi-period demographic evolution, or budget allocation. Accordingly, throughout this study, the term school network reorganization refers specifically to the aggregation and administrative dimensioning problem addressed by the proposed model.
We develop a synthetic benchmark instance generator to reproduce realistic school networks under different territorial and demographic scenarios, as no benchmark datasets currently exist for the school dimensioning problem. We also validate our approach on a real-world case study from the Calabria Region.
The proposed model is solved using Gurobi as a classical optimization approach, while we also investigate the applicability of Quantum Annealing (QA) to selected instances. QA has emerged as a promising paradigm for binary optimization problems characterized by large search spaces and complex constraint structures. The objective is to provide evidence of the model's suitability for QA rather than to outperform classical optimization methods.

The remainder of the paper is organized as follows. Section \ref{sec:Related works} reviews the main contributions in the literature on school network reorganization optimization approaches and recent developments in quantum optimization techniques for large-scale binary optimization problems.
Section \ref{sec:problem_definition} provides a detailed description of the problem context and its main dimensions, including administrative, territorial, and operational aspects. Section \ref{subsec:mathematical_formulation} presents two mathematical formulations of the school aggregation problem.
Section \ref{sec:computational_study} presents the computational study. It first describes the computational setup adopted for the experiments, then introduces the benchmark instance generation procedure, and finally reports and discusses the computational results obtained with the proposed optimization approaches. 
Section \ref{real} presents the real-world case study based on the school network of the Calabria region, describing the characteristics of the considered instance and reporting the computational results obtained through both classical and quantum optimization approaches.
Finally, Section \ref{sec:conclusions} concludes the paper and discusses possible directions for future research.

\section{Related works}\label{sec:Related works}

This section can be divided into two main research streams. The first concerns optimization models for school network planning. The second examines the application of quantum optimization techniques to combinatorial optimization problems.

\subsubsection{School network planning and aggregation works}
Early research on educational facility planning focused on determining the optimal number, location, and capacity of schools in response to demographic growth and changing spatial demand. One of the earliest optimization contributions is the model proposed by \citet{Obrien1969ModelFP}, which jointly considers school location, facility capacity, student allocation, and accessibility. This work laid the foundations for the application of operations research techniques to educational planning. As demographic dynamics became increasingly uncertain, researchers shifted toward long-term planning models. \citet{henig1986dynamic} developed a dynamic capacity planning model accounting for changing populations and decisions on school expansion, contraction, and reallocation, while \citet{greenleaf1987mathematical} proposed a mathematical programming model integrating school capacities, student assignments, and construction decisions. A complementary review by \citet{bruno1982analytical} highlighted the relevance of analytical approaches for educational facility planning under declining enrolment.

A parallel research stream addressed the school districting problem, focusing on the design of attendance boundaries under capacity, accessibility, and equity constraints. \citet{church1993modeling} proposed a multiobjective formulation for school consolidation that jointly considers school closures, student assignment, and capacity utilization. \citet{lemberg2000school} subsequently introduced temporal stability into school districting by minimizing the number of students affected by boundary changes, while \citet{caro2004school} incorporated GIS-based information to account for capacity, contiguity, geographic barriers, and travel distances.

Another line of research focused on strategic school network planning under changing demographic conditions. \citet{antunes2000dynamic} developed a multi-period facility location model jointly addressing school openings, closures, capacity expansion, and downsizing. Building on this perspective, \citet{teixeira2007optimization} proposed a $p$-median location-allocation model for the secondary school network of Coimbra, jointly determining school locations, capacities, and student assignments. Subsequent contributions incorporated additional aspects such as endogenous school choice~\citet{haase2013management}, GIS-based accessibility and attendance-zone redesign~\citet{alifi2017optimization}, and demand evolution, land availability, and budget constraints~\citet{barbara2021optimizing,dearmas2022improving}.

More recently, optimization research has increasingly focused on school consolidation in the context of demographic decline. \citet{bruno2016institutions} represent the closest contribution to the present work, addressing the Italian school dimensioning problem through the administrative consolidation of autonomous school institutions while preserving existing educational facilities. However, their formulation does not explicitly account for road-network accessibility, territorial fragility, or the broader regulatory framework governing school reorganization. Other recent studies have investigated the trade-offs associated with school consolidation, showing the importance of balancing accessibility, resource utilization, educational outcomes, and spatial equity~\citet{hannum2021estimating,bhatnagar2023sustainable,klein2024school,kim2025optimal}.

In this study, the regional school aggregation problem is addressed under territorial, institutional, and accessibility constraints. The proposed framework integrates institutional compatibility conditions, territorial criticality indicators, and real travel times computed on the road network within a unified binary optimization model. Furthermore, a decomposition-based optimization strategy is introduced to improve scalability for large-scale regional planning instances, separating provincial aggregation subproblems from the regional coordination phase.

A detailed comparison of the main optimization approaches is provided in Table~\ref{tab:literature_comparison}.
It summarizes the main characteristics of the school location and consolidation models reviewed in the literature. For each study, we report whether the mathematical formulation explicitly incorporates:
\textbf{Road-Network Travel Time}, accessibility measures based on road-network travel times, \textbf{Capacity}, school capacity or enrollment constraints, \textbf{School Aggregation}, school consolidation, merger, or aggregation decisions, 
\textbf{Institutional Constraints}, regulatory, administrative, or policy constraints governing school network design, 
\textbf{Territorial Fragility}, territorial vulnerability factors such as rurality, remoteness, demographic decline, or disadvantaged spatial conditions.

\begin{table}[h]
\centering
\caption{Comparison of optimization-based approaches for school network planning and aggregation. A \textbf{check} mark denotes explicit inclusion of the corresponding dimension in the objective function or constraints, \textbf{partial} indicates that the aspect is considered only indirectly or through simplified assumptions, whereas a \textbf{cross} denotes the absence of the corresponding modeling dimension.}
\label{tab:literature_comparison}
\resizebox{\columnwidth}{!}{
\begin{tabular}{
    L{4cm}
    C{2.2cm}
    C{2.2cm}
    C{2.3cm}
    C{2.8cm}
    C{2.2cm}
}
\hline
\textbf{Reference} &
\textbf{Road-Network Travel Time} &
\textbf{Capacity} &
\textbf{School Aggregation} &
\textbf{Institutional Constraints} &
\textbf{Territorial Fragility} \\
\hline

\citet{Obrien1969ModelFP} &
\xmark & \cmark & \xmark & \xmark & \xmark \\

\citet{bruno1982analytical} &
\xmark & \xmark & \xmark & \xmark & \xmark \\

\citet{henig1986dynamic} &
\xmark & \cmark & \xmark & \xmark & \xmark \\

\citet{greenleaf1987mathematical} &
\xmark & \cmark & \xmark & \xmark & \xmark \\

\citet{church1993modeling} &
\xmark & \cmark & \cmark & \xmark & \xmark \\

\citet{antunes2000dynamic} &
\xmark & \cmark & \xmark & \xmark & \xmark \\

\citet{lemberg2000school} &
\xmark & \cmark & \xmark & partial & \xmark \\

\citet{caro2004school} &
\cmark & \cmark & \xmark & partial & \xmark \\

\citet{teixeira2007optimization} &
\xmark & \cmark & \xmark & \xmark & \xmark \\

\citet{haase2013management} &
\xmark & \cmark & \xmark & \xmark & \xmark \\

\citet{bruno2016institutions} &
\xmark & \cmark & \cmark & partial & \xmark \\

\citet{alifi2017optimization} &
\cmark & \cmark & \xmark & \xmark & \xmark \\

\citet{barbara2021optimizing} &
\xmark & \cmark & \xmark & \xmark & \xmark \\

 \citet{hannum2021estimating} &
\xmark & \xmark & \cmark & \xmark & partial \\

\citet{dearmas2022improving} &
\xmark & \cmark & \xmark & \xmark & partial \\

\citet{bhatnagar2023sustainable} &
\cmark & \cmark & \cmark & \xmark & \xmark \\

\citet{klein2024school} &
\cmark & \xmark & \cmark & \xmark & \xmark \\

\citet{kim2025optimal} &
\xmark & \xmark & \cmark & \xmark & \cmark \\

This work &
\cmark & \cmark & \cmark & \cmark & \cmark \\

\hline
\end{tabular}
}

\end{table}

\subsubsection{Quantum Annealing for combinatorial optimization}

Recent research has increasingly investigated QA for combinatorial optimization, particularly through QUBO formulations and hybrid quantum-classical approaches~\citet{Glover2022QuantumBridgeI,Glover2022QuantumBridgeII}. Applications have emerged across several domains, including routing, facility location, packing, scheduling, logistics, and finance. A systematic overview is provided by \citet{ciacco2025review}, highlighting both the growing interest in quantum optimization and the technological limitations that currently affect its scalability. In particular, the transformation of constrained optimization models into QUBO formulations may require penalty terms, auxiliary variables, and problem-specific reformulations, while preprocessing can be used to reduce the resulting problem size before quantum execution.

Routing represents one of the most extensively investigated application areas. Early studies have explored QA for the Vehicle Routing Problem and related variants. \citet{tambunan2022quantum} investigated the VRP with Weighted Road Segments by formulating the problem as a QUBO and solving it on a D-Wave quantum annealer, while \citet{le2023quantum} addressed the Orienteering Problem through a QUBO formulation implemented on D-Wave systems. More complex logistics settings have also been considered. \citet{osaba2024solving} investigated delivery problems with heterogeneous vehicle fleets, delivery priorities, and weight and volume capacity constraints, using D-Wave's hybrid CQM solver on real-world instances. These contributions illustrate the increasing use of hybrid quantum-classical approaches for routing problems with realistic operational constraints.

Further research has focused on the Traveling Salesman Problem (TSP) and its variants, with particular attention to preprocessing and model-reduction strategies. Recent studies have investigated cost-based arc filtering, cutting-plane procedures, and network reduction to limit the size of the resulting quantum formulation~\citet{ciacco2026CAF, ciacco2025steiner}. In particular, the Steiner TSP has been extended to settings involving time windows and pickup-and-delivery constraints, showing how preprocessing and hybrid solution strategies can facilitate the treatment of richer routing models on current quantum hardware.

Facility location and network design are especially relevant to the present study. \citet{malviya2023logistics} investigated the strategic location of distribution centers through a QUBO-based hybrid quantum-classical approach, while \citet{ciacco2026facility} considered a two-level facility location problem and proposed a preprocessing procedure to reduce the size of the underlying network. These studies provide a direct methodological connection between QA and location-allocation decisions, which are central to school network planning.

In packing problems, quantum approaches have been investigated for different variants of the bin packing and related allocation problems, typically by reformulating the underlying constraints and objective functions into QUBO models suitable for quantum annealers~\citet{de2022hybrid,cellini2024qal,bozejko2024optimal,garcia2022comparative}. These studies highlight the importance of the QUBO formulation and penalty design in representing capacity and assignment constraints while keeping the number of binary variables manageable. Scheduling represents another relevant application domain, with QA approaches developed for job-shop scheduling and related personnel and resource allocation problems~\citet{venturelli2016job,perez2024solving,li2023quantum}. In these applications, quantum optimization is used to handle assignment, sequencing, and resource constraints, often within hybrid quantum-classical frameworks. Overall, these contributions show that QA can be applied to highly constrained combinatorial problems beyond routing and facility location, while also emphasizing the importance of suitable mathematical reformulations and problem-size reduction. In the educational context, \citet{ciacco2025Educational} applied QA to a staff allocation problem in a public school in Calabria, addressing the assignment of personnel across school sites and educational levels. Although this contribution is not concerned with strategic school network planning, it demonstrates the applicability of QA to constrained resource allocation problems within the educational sector.

To the best of our knowledge, no previous study has investigated QA for school aggregation or school dimensioning problems. The existing literature provides methodological evidence that QA can be applied to related routing, facility location, network design, and educational resource allocation problems, while the strategic planning and aggregation of school networks remain largely unexplored. This motivates the present investigation, which evaluates whether the proposed school network model can be effectively formulated and handled within a quantum optimization framework, using Gurobi as a classical reference.

\section{Problem Definition}
\label{sec:problem_definition}

The problem concerns the strategic reorganization of the school network through aggregation processes subject to complex regulatory, territorial, and operational constraints. The proposed framework is developed within the Italian school system, where the national Ministry of Education defines the legislative and regulatory framework, while Regional Governments are responsible for developing territorial plans for the organization of the school network. The optimization model therefore incorporates the institutional structure and policy rules governing school network planning in Italy.

A fundamental distinction is made between a physical school facility, i.e., the building where teaching activities take place, and an autonomous school institution, i.e., the administrative and legal entity responsible for its management. The aggregation process modeled in this study concerns exclusively the latter. Each aggregation process therefore involves two roles:
\begin{itemize}
\item the \textbf{school hub}, which retains its administrative autonomy and serves as the reference institution for the aggregated structure. The school hub is defined as the institution with the highest student enrollment among those involved in the aggregation;
\item the \textbf{aggregated school}, which loses its administrative autonomy and is incorporated into a school hub.
\end{itemize}

Importantly, aggregation does not imply the closure of the physical school facilities. Their activities continue to ensure the continuity of educational services, while administrative governance, staff management, and financial decision-making are transferred to the school hub and placed under a single principal.

The decision problem consists of partitioning a regional school network to determine which specific institutions retain their administrative autonomy, acting as surviving administrative hubs, and which lose their autonomy and are aggregated into other surviving structures. 

To fully capture the multi-faceted nature of the school aggregation process, the proposed optimization framework integrates six dimensions:

\begin{itemize}
    \item \textbf{Administrative geography and context}, which defines the institutional and territorial boundaries within which aggregation decisions can be made;
    
    \item \textbf{Institutional compatibility}, which determines whether schools can be aggregated according to their educational type and curricular structure;
    
    \item \textbf{Territorial fragility}, which captures the socioeconomic and infrastructural vulnerability of the areas where schools are located;
    
    \item \textbf{Differentiated dimensional requirements and eligibility}, which determine whether a school satisfies the enrollment thresholds required to retain autonomy;
    
    \item \textbf{Spatial accessibility}, which evaluates aggregation feasibility based on travel-time constraints over the road network;
    
    \item \textbf{Operational feasibility}, which ensures that the resulting aggregation remains sustainable in terms of capacity and organizational efficiency.
\end{itemize}

\subsection{Administrative geography and context}
The governance of public services is organized across a nested hierarchy of multiple administrative levels, typically involving national, regional, provincial, and local municipal authorities. 

This administrative structure directly imposes boundaries on the school aggregation process:
\begin{itemize}
    \item \textbf{Provincial Coherence:} due to governance requirements, school aggregations are strictly restricted within provincial borders; 
    
    \item \textbf{Cross-Municipal Penalization:} although schools located in different municipalities within the same province can legally be placed under a single administrative head, this configuration introduces significant logistical frictions and weakens local governance; 
\end{itemize}

\subsection{Institutional compatibility}

The Italian school system comprises different types of autonomous school institutions, characterized by distinct educational structures and administrative functions:

\begin{itemize}

\item \textbf{Comprehensive Institutes (CIs):}  
these represent the most common school structure and mainly serve younger students. They are widely distributed across the national territory and are present in most municipalities, ensuring local access to basic educational services. CIs typically include:
\begin{itemize}
    \item pre-primary education (ages 3--5), lasting 3 years;
    \item primary education (ages 6--10), lasting 5 years;
    \item lower secondary education (ages 11--13), lasting 3 years.
\end{itemize}

\item \textbf{Upper Secondary Institutes (USIs):}  
these institutions provide upper secondary education (ages 14--18), corresponding to a five-year educational cycle. Compared with CIs, they generally serve wider territorial areas and attract students from multiple municipalities. USIs include three main types of schools, each offering a distinct educational curriculum and covering different subject areas:
\begin{itemize}
    \item academic programs;
    \item technical programs;
    \item vocational programs.
\end{itemize}
    
\item \textbf{Integrated Institutes (II):}  
these institutions combine all educational stages within a single autonomous structure, from pre-primary to upper secondary education. They were mainly introduced in geographically isolated or low-density areas to ensure continuity in educational service provision, particularly in fragile territories. However, recent regulatory orientations have progressively discouraged the creation of new institutions of this type.

\item \textbf{Boarding School Institutions (BSI):}  
these represent a specific institutional category combining educational and residential functions. In addition to school activities, they provide accommodation and educational support services through dedicated educators. Their governance and organizational structure is typically more complex than that of standard school institutions.

\end{itemize}
Both IIs and BSIs represent highly specific and relatively rare institutional configurations within the Italian school system. Due to their particular regulatory and organizational characteristics, they are not considered in the dimensioning process analyzed in this study and are mentioned here only for completeness.

\begin{figure}[htbp]
    \centering
        \caption{Schematic representation of the main institutional configurations of the Italian school system, including educational levels and organizational structure.}
    \label{fig:school_system_1}
    \includegraphics[width=0.7\textwidth]{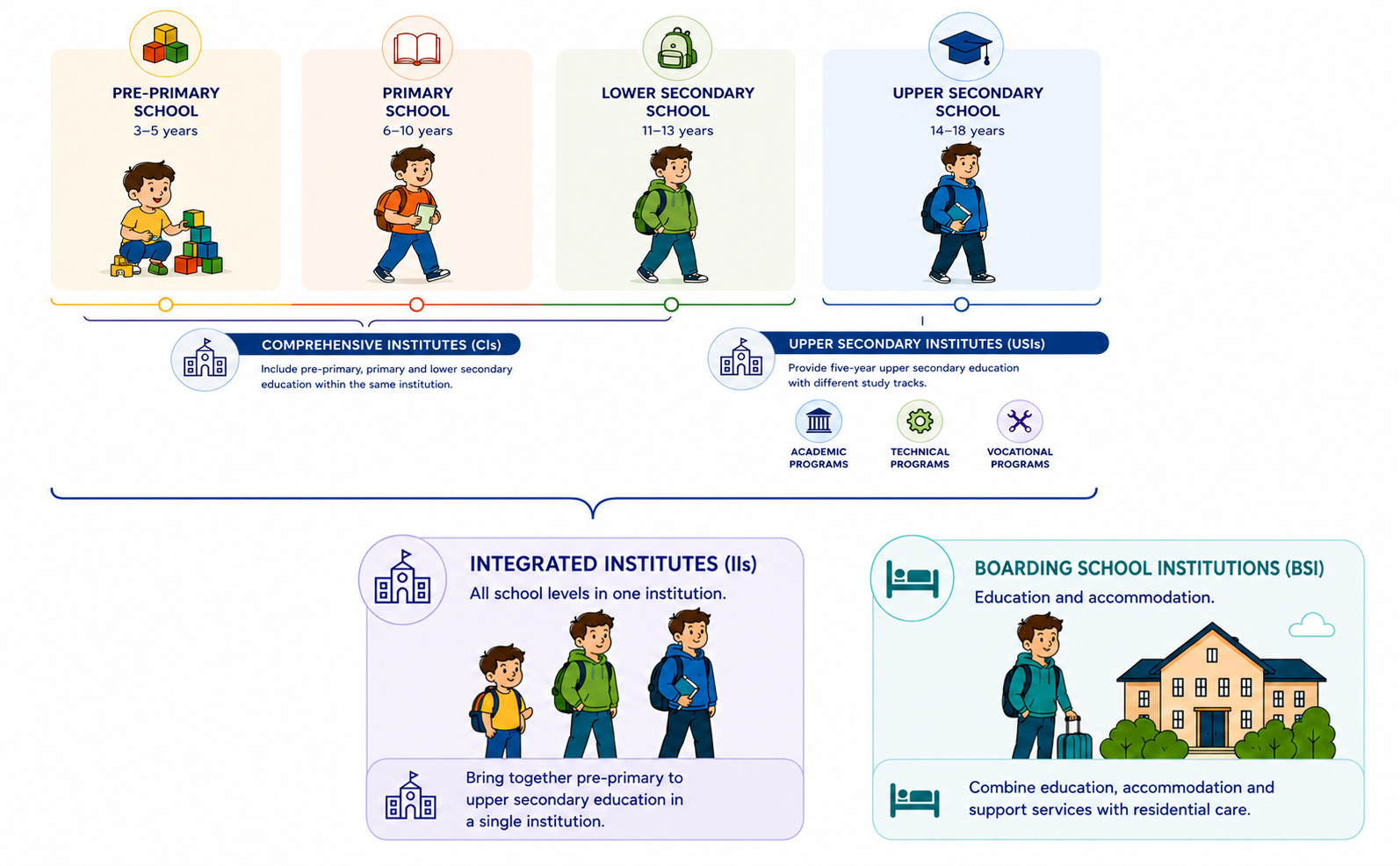}

\end{figure}

To facilitate the interpretation of the institutional structure described above, a schematic representation of the main school institution types considered in this study and their corresponding educational levels is provided in Figure \ref{fig:school_system_1}.

The institutional structure directly shapes the feasible aggregation choices within the optimization framework. In particular, school aggregations must satisfy compatibility conditions related to both educational level and curricular structure:
\begin{itemize}
\item \textbf{Educational-Level Compatibility:} schools can only be aggregated if they belong to the same educational category (CIs with CIs and USIs with USIs). Merging institutions with structurally different educational cycles is not allowed, both to preserve organizational coherence and to comply with the current Italian regulatory framework, which increasingly discourages the creation of integrated institutional structures such as IIs;

\item \textbf{Curricular Compatibility:} within the USIs, merging schools that offer different educational tracks is legally permitted but generates high internal management and organizational complexity. The framework accounts for this issue by modeling curricular heterogeneity. 
\end{itemize}

\subsection{Territorial Fragility}

School dimensioning raises a major socio-spatial challenge related to territorial fragility, defined as the structural vulnerability of a geographic area across demographic, socioeconomic, and infrastructural dimensions. In vulnerable contexts, a school may represent one of the last remaining forms of public infrastructure, serving as a pillar of community resilience and social cohesion.

To systematically incorporate this aspect, Italian regulations classify the municipality in which each school is located according to a territorial criticality index derived from national and regional planning guidelines. The classification evaluates four main dimensions of territorial vulnerability:
\begin{enumerate}
    \item \textbf{Income levels:} the proportion of low-income taxpayers within the municipality;
    \item \textbf{Education levels:} the percentage of higher-education graduates within the resident population;
    \item \textbf{Labor market:} the overall employment rate of the working-age population;
    \item \textbf{Institutional fragility:} structural vulnerabilities related to local municipal governance and public administration stability.
\end{enumerate}

By counting the cumulative number of critical indicators observed, the model establishes five increasing levels of territorial fragility, ranging from level zero (stable and well-served contexts without critical indicators) up to level four (severe socioeconomic vulnerability characterized by all four critical indicators). 

\subsection{Differentiated dimensional requirements and eligibility}
The territorial criticality level directly determines whether a school is eligible to participate in the dimensioning process or if it satisfies the structural requirements to potentially remain autonomous. Rather than enforcing a single, uniform size standard across the region, planning regulations establish asymmetric enrollment thresholds based on local vulnerability. This mechanism ensures that schools in fragile or isolated areas can preserve their administrative independence even with significantly lower numbers of enrolled students.

Operationally, the framework defines the set of eligible schools by validating each school's total student enrollment against differentiated statutory thresholds:
\begin{itemize}
    \item For schools in stable areas (level zero), the minimum enrollment required for automatic autonomy is set to the maximum standard threshold (one thousand students);
    
    \item For schools in low-criticality areas (level one), the required autonomy threshold drops to a slightly lower standard (eight hundred students).
    \item For schools in intermediate fragile areas (levels two and three), the threshold is further relaxed (six hundred students).
    \item For schools operating in the most vulnerable areas (level four), a highly protective threshold is applied (four hundred students), allowing very small institutions to safeguard their independent status.
\end{itemize}
Schools that do not reach the minimum size threshold, defined according to the vulnerability of their territory, are considered candidates for reorganization. Conversely, institutions that already exceed these protective thresholds or have a special institutional status are excluded from the dimensioning process by default and assumed to remain active.

\subsection{Spatial accessibility}
A viable school network must guarantee physical accessibility. Because regional geography is often highly complex, characterized by mountainous morphology, fragmented settlements, and heterogeneous transport networks, aggregation feasibility cannot be based on simple geometric distance. Instead, it must be evaluated using actual travel times measured over the real road transportation network.

To avoid imposing unmanageable travel burdens on students and administrative staff, two institutions can be aggregated only if the travel time between them does not exceed a maximum feasible threshold. Crucially, these limits are differentiated by educational level to respect different mobility patterns and age groups:
\begin{itemize}
    \item \textbf{CIs:} because these institutes serve younger children, they require strong proximity constraints and a strict, low maximum travel time.
    \item \textbf{USIs:} since these institutes serve older, more mobile students, they are assigned a wider maximum travel-time threshold, allowing them to draw from larger territorial catchment areas.
\end{itemize}

\subsection{Operational feasibility}
To guarantee that the resulting network configuration is managerially sustainable and organizationally realistic, the framework enforces additional operational dimensions:
\begin{itemize}
    \item \textbf{Balanced Aggregation Principle:} to maintain administrative stability, smaller and structurally weaker schools are consolidated into institutions that possess an equal or larger student enrollment base. An aggregation flow can only move from a smaller school toward a larger or equal-sized peer.
    \item \textbf{Capacity Limits:} the total cumulative number of students assigned to a surviving active institution (including its own students plus all students from absorbed schools) cannot exceed predefined capacity thresholds, preventing the creation of unmanageable, oversized mega-institutions.
    \item \textbf{Global Reorganization Target:} the overall scope of the dimensioning policy is dictated by a macro-level parameter, which defines the exact minimum proportion of schools that must remain active as autonomous entities after the optimization process.
\end{itemize}

\section{Mathematical Formulation} \label{subsec:mathematical_formulation}
This section introduces the notation and definitions required for the mathematical formulations of the problem and presents the corresponding mathematical models.

Let $\mathcal{N}$ denote the set of all schools in the region, with cardinality $n=|\mathcal{N}|$. 
Let $N\subseteq\mathcal{N}$ be the subset of schools that are eligible for the dimensioning process, while the remaining $\bar n=n-|N|$ schools are assumed to remain active and therefore do not participate in the aggregation decisions. Let $N^{CI}\subseteq N$ denote the subset of CI, and let $N^{USI}\subseteq N$ denote the subset of USI. 
For each pair of schools $(i,j)\in N\times N$, let $t_{ij}$ denote the travel time between schools $i$ and $j$, computed on the road transportation network and expressed in seconds.

The binary parameter $\delta_{ij}$ is equal to $1$ if schools $i$ and $j$ belong to the same educational category (CI or USI), and $0$ otherwise. The binary parameter $\theta_{ij}$ is equal to $1$ if two USI offer different educational tracks, and $0$ otherwise. Hence, $\delta_{ij}$ enforces compatibility at the institutional level, whereas $\theta_{ij}$ captures curricular incompatibilities within the upper secondary education system.

The binary parameter $\alpha_{ij}$ is equal to $1$ when schools $i$ and $j$ are located in the same province, and $0$ otherwise.
The binary parameter $\beta_{ij}$ is equal to $1$ when schools $i$ and $j$ belong to different municipalities, and $0$ otherwise.

Each school $i$ is associated with a number of enrolled students $s_i$ and a territorial criticality level $g_i$, derived from the characteristics of the municipality in which the school is located. Based on this classification, we define two subsets, i.e., $N_3$ and $N_4$, representing schools located in municipalities with criticality levels 3 and 4, respectively.

Let $T^{CI}_{\max}$ denote the maximum feasible travel time for aggregations involving CI, and let $T^{USI}_{\max}$ denote the corresponding threshold for USI. Different thresholds are adopted to reflect the distinct territorial catchment areas and mobility patterns associated with the two educational levels.
Let 
$A=
\Big\{
(i,j)\in N\times N :
i\neq j,\;
s_i\le s_j,\;
\big[
(i\in N^{CI} \wedge t_{ij}\le T_{\max}^{CI})
\vee
(i\in N^{USI} \wedge t_{ij}\le T_{\max}^{USI})
\big]
\Big\}
$
denote the set of candidate aggregation arcs. 
An arc $(i,j)\in A$ indicates that school $i$ can potentially be aggregated into school $j$, provided that school $j$ has at least the same enrollment as school $i$ and that the travel time between the two schools satisfies the feasible threshold associated with the educational category of school $i$.


The parameter $\gamma \in (0,1)$ defines the proportion of schools that must remain active after the reorganization process.

The parameters used in the formulation of the problem are summarized in Table \ref{tab:notations}.
\begin{table}[h]
\centering
\caption{Summary of the notation used in the mathematical formulation of the optimization model.}
\label{tab:notations}
\begin{tabular}{>{\centering\arraybackslash}m{3cm}>{\arraybackslash}m{13cm}}
\toprule
\textbf{Notation} & \textbf{Description} \\
\midrule
$\mathcal{N}= \{1,\dots,n\}$ & Set of all schools in the territory \\

$n$ & Total number of schools in the region \\

$N \subseteq \mathcal{N}$ & Set of schools eligible for the dimensioning process \\

$N^{CI}$ & Set of CIs \\

$N^{USI}$ & Set of USIs \\
$N_3$ & Set of schools located in municipalities with criticality level 3 \\

$N_4$ & Set of schools located in municipalities with criticality level 4 \\
$\bar{n}$ & Number of schools not involved in the dimensioning process \\

$A$ & Set of candidate aggregation arcs \\

\hline
$t_{ij}$ & Travel time between schools $i$ and $j$ measured on the road network \\

$\delta_{ij}$ & 1 if schools $i$ and $j$ belong to the same educational category (CI or USI), 0 otherwise \\

$\alpha_{ij}$ & 1 if schools $i$ and $j$ are located in the same province, 0 otherwise \\

$\beta_{ij}$ & 1 if schools $i$ and $j$ belong to different municipalities, 0 otherwise \\

$\theta_{ij}$ & 1 if two upper secondary schools offer different educational tracks, 0 otherwise \\

$s_i$ & Number of students enrolled in school $i$ \\

$g_i$ & Territorial criticality level of school $i$ \\

$\bar{z}$ & Maximum number of schools that can remain active after aggregation \\

$T_{\max}^{CI}$ & Maximum admissible travel time for aggregations involving CIs \\

$T_{\max}^{USI}$ & Maximum admissible travel time for aggregations involving USIs \\

$\gamma$ & Minimum proportion of schools that must remain active after reorganization \\

\bottomrule
\end{tabular}

\end{table}

Two alternative formulations of the school aggregation problem are proposed, progressively refined to improve compactness. The first serves as a baseline integer programming model, in which compatibility conditions are enforced through explicit constraints. The second reformulates the problem as a capacitated facility location problem (CFLP), embedding compatibility directly into the domain of the assignment variables to reduce the constraint structure. A formal equivalence between the two formulations is then established, showing that they provide an equivalent representation of the school aggregation problem and admit the same feasible solutions and optimal objective value.

\subsection{Variant I: Baseline integer programming formulation}\label{sec:variant1}

The model is defined by the following binary decision variables:
\begin{description}
    \item[$z_i$] Binary variable, $\forall i\in N$:
    $\begin{cases}
    1 & \text{if school $i$ remains active (i.e., it is selected as a school hub)},\\
    0 & \text{otherwise.}
    \end{cases}$
\end{description}

\begin{description}
        \item[$y_{ij}$] Binary variable, $\forall (i,j)\in A$:
    $\begin{cases}
    1 & \text{if school $i$ is assigned (aggregated) to school $j$,}\\
    0 & \text{otherwise.}
    \end{cases}$
\end{description}

The mathematical formulation can be written as follows:
\begin{align}
\min \quad
&
C_1 \sum_{i \in N}
\sum_{\substack{j \in N \\ (i,j) \in A}}
\beta_{ij} y_{ij}
+
C_2 \sum_{i \in N}
\sum_{\substack{j \in N \\ (i,j) \in A}}
\theta_{ij} y_{ij}
\nonumber\\
&
\quad
+
C_3 \sum_{i \in N_3}
\sum_{\substack{j \in N \\ (i,j) \in A}}
y_{ij}
+
C_4 \sum_{i \in N_4}
\sum_{\substack{j \in N \\ (i,j) \in A}}
y_{ij}
\nonumber\\
&
\quad
+
\sum_{i \in N}
\sum_{\substack{j \in N \\ (i,j) \in A}}
\left(
C_5 (1-\beta_{ij})
+
C_6 (1-\theta_{ij})
\right)
y_{ij}
\label{obj:v1}
\\[0.5em]
\text{s.t.}\qquad
&
z_i
+
\sum_{j \in N:\,(i,j)\in A}
y_{ij}
= 1,
&&
\forall i \in N
\label{assign:v1}
\\
&
y_{ij}
\le z_j,
&&
\forall (i,j)\in A
\label{activation}
\\
&
2y_{ij}
\le
\delta_{ij}
+
\alpha_{ij},
&&
\forall (i,j)\in A
\label{compatibility}
\\
&
\bar{n}
+
\sum_{i \in N} z_i
\le
\gamma\, n
\label{numschools}
\\
&
s_j
+
\sum_{i \in N:\,(i,j)\in A}
s_i y_{ij}
\le
1500,
&&
\forall j \in N
\label{capacity:v1}
\\
&
z_i \in \{0,1\},
&&
\forall i \in N
\label{eq:binary:v1}
\\
&
y_{ij} \in \{0,1\},
&&
\forall (i,j)\in A
\label{eq:binary1:v1}
\end{align}
The objective function \eqref{obj:v1} consists of multiple components designed to capture the main policy objectives of the school aggregation process. The first term penalizes aggregations between schools located in different municipalities, thereby discouraging cross-municipal consolidations that may increase coordination complexity and weaken local governance. The second term penalizes aggregations between upper secondary schools with different curricular tracks, reflecting the additional organizational complexity associated with heterogeneous educational structures. The third and fourth terms introduce penalties for aggregations involving schools located in municipalities with territorial criticality levels 3 and 4, respectively. These terms aim to discourage solutions that may further reduce institutional autonomy in fragile and socioeconomically vulnerable areas. The fifth term rewards aggregations between schools belonging to the same municipality, thus favoring geographically and administratively coherent consolidations. The sixth term rewards aggregations between schools with compatible curricular structures, promoting solutions that reduce organizational complexity and improve managerial efficiency.

Constraints \eqref{assign:v1} ensure that each school is assigned to exactly one configuration: either it remains active as an autonomous institution (when $z_i = 1$) or it is aggregated into another school (when $y_{ij}=1$ for some $j \in N$). 
Constraints \eqref{activation} impose activation consistency by ensuring that a school can receive aggregated institutions only if it remains active. Therefore, only selected school hubs can act as aggregation centers.
Constraints \eqref{compatibility} enforce structural compatibility between aggregated schools. In particular, an aggregation is allowed only if the two schools belong to the same educational category and they are located within the same province.
Constraint \eqref{numschools} regulates the global reorganization target by limiting the number of schools that remain active after the aggregation process according to the parameter $\gamma$.
Constraints \eqref{capacity:v1} impose capacity restrictions on active schools by ensuring that the total number of enrolled students assigned to each school hub, including both its own students and those of aggregated schools, does not exceed the maximum admissible threshold.
Finally, constraints \eqref{eq:binary:v1} -\eqref{eq:binary1:v1}  define the binary nature of the decision variables.


\subsection{Variant II: Capacitated facility location formulation}\label{sec:variant2}



We introduce an alternative optimization model based on the CFLP. This formulation is mathematically equivalent to the \ref{sec:variant1} formulation presented above, but it provides a more compact representation by incorporating compatibility constraints directly into the definition of the feasible variable domains.
In this CFLP-based formulation, facilities are assumed not to incur opening costs and can only be activated at demand locations corresponding to eligible schools. Accordingly, we redefine the operational domain and the associated parameters to capture the specific features of the school reorganization process.

In particular, let $\tilde{A}$ denote the restricted set of feasible aggregation arcs, which incorporates both institutional category compatibility requirements and provincial boundary restrictions, as defined below:
$
\tilde{A}=
\Big\{
(i,j)\in N\times N :
i\neq j,\;
s_i\le s_j,\;
\big[
(i\in N^{CI} \wedge t_{ij}\le T_{\max}^{CI})
\vee
(i\in N^{USI} \wedge t_{ij}\le T_{\max}^{USI})
\big],\;
\delta_{ij}+\alpha_{ij}=2
\Big\}.
$
Thus, $\tilde{A}\subseteq A$ includes only those arcs satisfying both operational feasibility and institutional compatibility, namely schools belonging to the same educational category and located within the same province.

Furthermore, for each school $j \in N$, we define the residual student capacity as $c_j = 1500 - s_j$, where $s_j$ denotes the current student enrollment of school $j$. Let $\bar{z}=(\gamma n)-\bar{n}$ denote the maximum number of schools that can remain active after the aggregation process with $\gamma \in (0,1)$.

The binary variable \(y_{ij}\), $\forall (i,j) \in \tilde{A}$ is defined as: 
\begin{description}
\centering
    \item[$y_{ij}$] $\begin{cases}

1 & \text{if school } i \text{ is assigned (aggregated) to school } j;\\
0 & \text{otherwise}.
\end{cases}$
\end{description}
The resulting mathematical formulation is reported in the following:

\begin{align}
\min \quad
&
C_1 \sum_{i \in N}
\sum_{\substack{j \in N \\ (i,j) \in \tilde{A}}}
\beta_{ij} y_{ij}
+
C_2 \sum_{i \in N}
\sum_{\substack{j \in N \\ (i,j) \in \tilde{A}}}
\theta_{ij} y_{ij}
\nonumber\\
&
\quad
+
C_3 \sum_{i \in N_3}
\sum_{\substack{j \in N \\ (i,j) \in \tilde{A}}}
y_{ij}
+
C_4 \sum_{i \in N_4}
\sum_{\substack{j \in N \\ (i,j) \in \tilde{A}}}
y_{ij}
\nonumber\\
&
\quad
+
\sum_{i \in N}
\sum_{\substack{j \in N \\ (i,j) \in \tilde{A}}}
\left(
C_5 (1-\beta_{ij})
+
C_6 (1-\theta_{ij})
\right) y_{ij}
\label{obj:v2}
\\[0.5em]
\text{s.t.}\qquad
&
z_i
+
\sum_{j \in N:\,(i,j)\in\tilde{A}} y_{ij}
= 1,
&&
\forall i \in N
\label{assign:v2}
\\
&
\sum_{i \in N:\,(i,j)\in\tilde{A}} s_i y_{ij}
\le c_j z_j,
&&
\forall j \in N
\label{capacity:v2}
\\
&
\sum_{i \in N} z_i
\le \bar{z}
\label{numschools:v2}
\\
&
z_i \in \{0,1\},
&&
\forall i \in N
\label{eq:binary:v2}
\\
&
y_{ij} \in \{0,1\},
&&
\forall (i,j)\in\tilde{A}
\label{eq:binary1:v2}
\end{align}

The objective function \eqref{obj:v2} follows the same rationale as the objective function \eqref{obj:v1} introduced in Variant \ref{sec:variant1} and preserves the same policy objectives.
Constraints \eqref{assign:v2} retains the same meaning and role as constraints \eqref{assign:v1}. 
Constraints \eqref{capacity:v2} represent the capacity restrictions of activated school hubs. Specifically, the total number of students assigned to a school hub cannot exceed its residual capacity $c_j$, thereby ensuring operational feasibility and preventing oversized institutional configurations.
Constraint \eqref{numschools:v2} regulates the global reorganization target by limiting the number of schools that remain active after the consolidation process.
Finally, constraints \eqref{eq:binary:v2} and \eqref{eq:binary1:v2} define the binary nature of the decision variables.

\subsection{Equivalence of the two formulations}
\label{sec:equivalence}

Variant I and Variant II represent the same school aggregation problem, but they describe feasibility in two different ways. Variant I starts from the complete set of candidate aggregation arcs $A$ and explicitly imposes compatibility, hub activation, and capacity constraints. Variant II removes incompatible arcs in advance by defining the restricted set $\widetilde{A}$ and combines the activation and capacity requirements into a single constraint.

The purpose of this section is to prove that the two formulations preserve the same feasible solutions and objective value.

\subsubsection{Compatibility and feasible arcs}

In Variant I, compatibility is enforced by constraint~\eqref{compatibility}, whereas Variant II embeds the same requirements directly in the definition of $\widetilde{A}$.

Consider an arc $(i,j)\in A\setminus\widetilde{A}$. By definition of $\widetilde{A}$, at least one of the two compatibility conditions is not satisfied, so $\delta_{ij}+\alpha_{ij}\leq1$. Constraint~\eqref{compatibility} therefore forces $y_{ij}=0$. Hence, all variables associated with arcs in $A\setminus\widetilde{A}$ are fixed to zero in every feasible solution of Variant I.

Therefore, these variables do not affect the assignment decisions and can be removed from the model without eliminating any feasible solution. Consequently, constraint~\eqref{assign:v1} can be restricted to $\widetilde{A}$, resulting in constraint~\eqref{assign:v2}, without changing the feasible aggregation configurations.

Conversely, consider any feasible solution of Variant II. By setting $y_{ij}=0$ for all $(i,j)\in A\setminus\widetilde{A}$, the solution can be extended to the original arc set $A$. The compatibility condition in constraint~\eqref{compatibility} is then satisfied, while all assignment decisions on $\widetilde{A}$ remain unchanged. Thus, every feasible solution of Variant II corresponds to a feasible solution of Variant I.

Hence, restricting the assignment variables from $A$ to $\widetilde{A}$ does not modify the set of feasible aggregation configurations.

\subsubsection{Activation and capacity}

In Variant I, activation and capacity are enforced separately by constraints~\eqref{activation} and~\eqref{capacity:v1}. Variant II replaces these two constraints with the single constraint~\eqref{capacity:v2}.

To prove equivalence, consider first a feasible solution of Variant I. If $z_j=0$, constraint~\eqref{activation} prevents any school from being assigned to $j$, so the left-hand side of constraint~\eqref{capacity:v2} is zero. If $z_j=1$, constraint~\eqref{capacity:v1} directly ensures that the assigned students do not exceed the residual capacity $c_j$. Therefore, in both cases, constraint~\eqref{capacity:v2} is satisfied.

Conversely, consider a solution satisfying constraint~\eqref{capacity:v2}. If $z_j=0$, the right-hand side is zero. Since $s_i>0$ and $y_{ij}$ is binary, no school can be assigned to $j$, and the activation condition is therefore satisfied. If $z_j=1$, constraint~\eqref{capacity:v2} reduces directly to the residual-capacity condition imposed by constraint~\eqref{capacity:v1}. Hence, constraint~\eqref{capacity:v2} is equivalent to the combination of the activation and capacity constraints of Variant I.

\subsubsection{Global reorganization constraint}

The two formulations impose the same requirement on the maximum number of schools that can remain administratively active.

Variant I imposes constraint~\eqref{numschools}, while Variant II uses constraint~\eqref{numschools:v2}, where $\bar{z}$ is defined as $\bar{z}=\gamma n-\bar{n}$.

Substituting this definition into constraint~\eqref{numschools:v2} gives

\begin{equation}
\sum_{i\in N}z_i
\leq \gamma n-\bar{n}.
\end{equation}

Adding $\bar{n}$ to both sides yields constraint~\eqref{numschools}. Therefore, the two formulations impose exactly the same upper bound on the number of schools that remain active after the reorganization.

\subsubsection{Objective function}

The two formulations use the same objective function, differing only in the set of arcs over which the assignment variables are defined: $A$ in Variant I and $\widetilde{A}$ in Variant II.

Consider a feasible solution of Variant I. From the compatibility analysis above, all variables associated with arcs in $A\setminus\widetilde{A}$ are equal to zero. Therefore, these arcs contribute nothing to the objective function, and restricting the objective from $A$ to $\widetilde{A}$ does not change its value.

Conversely, consider a feasible solution of Variant II and extend it to $A$ by setting $y_{ij}=0$ for all $(i,j)\in A\setminus\widetilde{A}$. Since the additional variables are zero, they do not contribute to the objective function, and the objective value remains unchanged.

Hence, the correspondence between the two formulations preserves the objective value for every pair of corresponding feasible solutions.

\subsubsection{Equivalence result}

The previous arguments establish a one-to-one correspondence between the feasible solutions of Variant I and Variant II. A feasible solution of Variant I can be restricted to $\widetilde{A}$ by removing variables that are necessarily equal to zero, while a feasible solution of Variant II can be extended to $A$ by assigning zero to all removed variables.

This correspondence preserves the assignment decisions, the set of active school hubs, compatibility, capacity feasibility, the global reorganization requirement, and the objective value.

Therefore, the two formulations define the same optimization problem and have the same optimal objective value:

\begin{equation}
\mathrm{OPT}_{\mathrm{V1}}
=
\mathrm{OPT}_{\mathrm{V2}}.
\label{eq:equivalence_result}
\end{equation}

Variant II is consequently an exact compact reformulation of Variant I. Its advantage is not a change in the underlying optimization problem, but a more compact representation in which incompatible aggregation arcs are removed before optimization and the activation and capacity requirements are represented by a single constraint.

\section{Computational Study}
\label{sec:computational_study}

This section describes the computational experiments conducted to assess the performance of the proposed optimization approaches for the school aggregation problem. The empirical study is organized into three distinct parts. First, Section~\ref{sec:computational_setup} outlines the computational setup, specifying the hardware environment, software libraries, and the algorithmic configurations adopted for both classical and quantum solvers. Second, Section~\ref{sec:instance_generation} details the instance generation procedure used to construct benchmark instances of varying sizes and territorial complexities based on real-world demographic data. Finally, Section~\ref{sec:numerical_results} reports and analyzes the numerical results obtained by solving the proposed Variant \ref{sec:variant2} formulation across all tested configurations, providing a comparative performance evaluation in terms of solution quality, feasibility, and computational runtime.

\subsection{Computational Setup}\label{sec:computational_setup}

This section outlines the computational environment and the algorithmic configurations adopted for the experimental evaluation. To assess the effectiveness of the proposed methodology for the school aggregation problem, the experimental framework compares two solving strategies:
\begin{enumerate}
    \item \textbf{Classical Optimization:} exact solution with Gurobi Optimizer, used as a benchmark for comparison with the solutions obtained with hybrid quantum-classical approaches.
    \item \textbf{Hybrid Quantum-Classical Optimization:} solution through D-Wave's cloud-based hybrid framework, 
aimed at evaluating the applicability of hybrid quantum-classical optimization to the considered constrained binary problem.

\end{enumerate}

All algorithms were implemented in Python and the computational campaigns were executed on a workstation equipped with an Intel Core processor, 32~GB RAM, and a 64-bit operating system.

\subsubsection{Classical Optimization Setup}

For the classical computation we use Gurobi Optimizer version 11.0.1 \citet{gurobi-doc}. Formulated as a constrained binary programming problem, the school aggregation model is characterized by a high-dimensional decision space bounded by dense institutional, territorial, and accessibility constraints. Consequently, the resulting formulation falls within the class of NP-hard combinatorial optimization problems. 
Gurobi serves as the exact optimization benchmark to evaluate the solution quality of the quantum approaches. Computational performance is quantitatively measured based on objective function values, computational runtime, and solution feasibility.

\subsubsection{Hybrid Quantum-Classical Setup}

A hybrid quantum-classical approach combines quantum computing with classical optimization techniques, allowing the two computational paradigms to work together within the same optimization workflow. In this setting, the quantum processor is not used to solve the entire optimization problem independently; instead, quantum and classical computational resources are coordinated to search for high-quality solutions to the constrained optimization problem.
A hybrid quantum-classical pipeline was deployed using D-Wave’s Leap cloud platform. The model was formalised as a CQM using \texttt{dimod} (v0.12.18) and \texttt{dwave-system} (v1.28.0), and submitted to the \texttt{hybrid\_constrained\_quadratic\_model\_version1p} service via the \texttt{LeapHybridCQMSampler}.

The hybrid quantum-classical solver is not an exact optimization method and does not guarantee the identification of the global optimum within a single run. Its search procedure combines quantum and classical heuristic components and is therefore probabilistic, meaning that repeated executions of the same model may explore different regions of the solution space and return different feasible solutions. To account for this stochastic behaviour, each test instance is solved independently five times under the same model and parameter settings. Repeating each experiment allows us to assess the robustness and variability of the obtained solutions and reduces the influence of a single stochastic run on the experimental results. The results reported for each instance are therefore evaluated across the five independent runs rather than being based on a single execution.

For a broader discussion on hybrid quantum-classical architectures and their applications in combinatorial optimization, see \citet{ciacco2026cutting, bertuzzi2024evaluation, ciacco2025steinerv2}.

The hybrid paradigm is particularly suitable for the school aggregation problem because it allows the constrained structure of the problem to be preserved. While a QUBO formulation requires the original constraints to be incorporated into the objective function through penalty terms~\citep{Khan2025Nash}, the CQM formulation adopted in this study represents them explicitly. This provides a more direct representation of the institutional, geographical, and capacity requirements while leveraging the hybrid quantum-classical capabilities of the D-Wave framework~\citep{DWaveCQMdocs}.

\subsection{Instance Generation}
\label{sec:instance_generation}

Since no publicly available benchmark dataset exists for the school aggregation problem under the regulatory and territorial constraints, a structured random instance generator was developed to produce synthetic yet realistic test cases for evaluating the scalability of the proposed model across different problem sizes. The generator reproduces the key structural features of a regional school network, including the hierarchical settlement geography, the spatial distribution of institutions, the heterogeneity of school types and educational tracks, and the differentiated territorial criticality levels that characterize fragile regional contexts. 
Four instance sizes were considered with $n \in \{250, 500, 750, 1000\}$, spanning from medium-scale to large-scale regional planning scenarios. 

\subsubsection{Spatial Domain and Coordinate Generation}
\label{subsubsec}

The synthetic territory is modeled as a square planar domain of side length proportional to the instance size, corresponding to the spatial region $[0,n]\times[0,n]$, where $n$ denotes the number of schools in the instance and coordinates are expressed in abstract spatial units. This design ensures that the spatial extent of the territory grows proportionally with the instance size, preserving an approximately constant average school density across instances of different scales.

To reproduce a realistic territorial organization, the synthetic territory is structured hierarchically into provinces and municipalities. Municipalities are divided into four categories representing different settlement scales: provincial capitals, large municipalities, medium municipalities, and small municipalities. This hierarchical design reflects the heterogeneous spatial organization of real-world territories, where large urban centers coexist with medium settlements and small peripheral communities.

The number of municipalities is determined as a function of the instance size. Specifically, the cardinality of each municipality category is defined as:

\begin{equation}
n_{\text{cap}}=\max(1,\lfloor 0.01n\rfloor), \quad
n_{\text{large}}=\left\lfloor \frac{n}{6}\right\rfloor,\quad
n_{\text{med}}=\left\lfloor \frac{n}{4}\right\rfloor,\quad
n_{\text{small}}=\left\lfloor \frac{n}{2}\right\rfloor.
\end{equation}

The total number of municipalities is therefore:

\begin{equation}
M=n_{\text{cap}}+n_{\text{large}}+n_{\text{med}}+n_{\text{small}}.
\end{equation}

This scaling mechanism ensures that larger instances generate progressively richer and more articulated territorial structures.

Each municipality type is characterized by two structural attributes: a spatial dispersion radius (i.e., $r$) and a territorial criticality level (i.e., $g$). Provincial capitals correspond to large urban centers with the broadest spatial footprint and the lowest criticality. Large municipalities represent medium-to-large urban settlements with moderate vulnerability. Medium municipalities capture semi-peripheral communities with heterogeneous criticality levels, while small municipalities represent peripheral and fragile areas characterized by compact geographic structure and higher criticality.

In accordance with the Italian regulatory framework, schools located in small municipalities are exclusively assigned to the CI category, reflecting the concentration of essential educational services in peripheral areas.

Schools are assigned to municipalities through a weighted random allocation process. Each municipality type is associated with a predefined allocation weight, determining the probability that a school is assigned to municipalities of that category. Higher weights are assigned to urban municipalities, reflecting their larger educational capacity and higher concentration of schools.

The structural characteristics of the municipality categories adopted in the synthetic instance generator are summarized in \ref{app:instance_params} (Table \ref{tab:municipality_types}).

Each municipality $m \in \{1,\dots,M\}$ is assigned a geographic center $(\bar{\text{lat}}_m,\bar{\text{lon}}_m)$ sampled independently and uniformly over the spatial domain: $(\bar{\text{lat}}_m,\bar{\text{lon}}_m)\sim U([0,n]\times[0,n]).
$

Once a school $i$ is assigned to municipality $m$, its geographic coordinates are generated according to a bivariate Gaussian distribution centered at the municipality center: $(\text{lat}_i,\text{lon}_i)\sim
\mathcal{N}
\left(
(\bar{\text{lat}}_m,\bar{\text{lon}}_m),
r_m^2\mathbf{I}_2
\right), $ where $r_m$ denotes the municipality-specific dispersion radius and $\mathbf{I}_2$ is the $2\times2$ identity matrix. Coordinates falling outside the spatial domain are clipped to the interval $[0,n]$.

This generation mechanism naturally produces spatial clusters of schools around municipality centers. Larger municipalities generate wider and more dispersed clusters, while smaller municipalities produce denser and more localized distributions. As a result, the synthetic instances exhibit heterogeneous geographic patterns that realistically mimic the coexistence of urban concentration and territorial fragmentation.

After generating the synthetic territorial structure, schools are assigned to municipalities through a weighted random allocation procedure. Each municipality $m$ is associated with an allocation weight $w_m$, determined by its category and proportional to its spatial scale. This reflects the assumption that larger and more urbanized municipalities tend to host more schools than smaller peripheral communities.

Formally, the probability of assigning a school to municipality $m$ is defined as:

\begin{equation}
    \mathbb{P}(\text{school assigned to } m) = \frac{w_m}{\sum_{m' \in \mathcal{M}} w_{m'}};
    \label{eq:allocation_prob}
\end{equation}

\noindent where $\mathcal{M}$ denotes the set of all municipalities and $w_m$ represents the allocation weight corresponding to the category of municipality $m$,
as reported in Table \ref{tab:municipality_types}.

Municipalities belonging to larger urban categories are assigned higher weights, resulting in a higher probability of receiving schools. Conversely, small municipalities receive lower weights and therefore host fewer schools on average. This allocation mechanism ensures that the synthetic instances reproduce realistic territorial patterns, where schools are concentrated in urban areas while remaining sparsely distributed across peripheral and fragile regions.

\subsubsection{Travel Time Computation}
\label{subsubsec:travel_times}

Travel times $t_{ij}$ between schools $i$ and $j$ are computed from the Euclidean distance between their generated coordinates: $d_{ij} = \sqrt{(\text{lat}_i - \text{lat}_j)^2 + (\text{lon}_i - \text{lon}_j)^2}.$
    
\noindent Travel time is then obtained by dividing the Euclidean distance by a fixed average speed parameter $v > 0$: $ t_{ij} = \frac{d_{ij}}{v}.$

In all experiments, we set $v = 1$, so that $t_{ij} = d_{ij}$ in spatial units. This assumption is consistent with the synthetic nature of the instances, where the spatial domain is abstract and no conversion to real-world units is required.

The travel-time thresholds $T_{\max}^{CI}$ and $T_{\max}^{USI}$ are computed dynamically from the road-network data as convex combinations of the minimum and maximum travel times observed across all compatible school pairs:
\begin{equation}
T_{\max}^{CI}
=
\eta_{CI} \, t_{\max}
+
(1-\eta_{CI}) \, t_{\min},
\qquad
T_{\max}^{USI}
=
\eta_{USI} \, t_{\max}
+
(1-\eta_{USI}) \, t_{\min};
\label{convex}
\end{equation}
where $\eta_{CI}, \eta_{USI} \in (0,1)$ are tunable parameters, and $t_{\min}$ and $t_{\max}$ denote the minimum and maximum travel times, respectively, among all compatible school pairs.

\subsubsection{Eligibility Filtering}
\label{subsubsec:eligibility}

Once all school attributes have been generated, the eligibility filtering procedure described in Section~\ref{sec:problem_definition} is applied to partition the full school set $\mathcal{N}$ into the eligible subset $N$ and the set of schools that retain their autonomy. Specifically, school $i$ is considered eligible for the dimensioning process if its enrollment $s_i$ does not exceed the criticality-dependent threshold associated with its territorial criticality level $g_i$: 1000 students for $g_i=0$, 800 for $g_i=1$, 600 for $g_i\in\{2,3\}$, and 400 for $g_i=4$.

Schools exceeding these thresholds are excluded from the optimization model and are assumed to remain active throughout the 
planning horizon. The number of excluded schools, denoted by $\bar{n}$, varies across instances depending on the realized enrollment distribution and the criticality composition of the generated territory.

\subsubsection{School Attributes}
\label{subsubsec:school_attributes}

Each school $i$ is characterized by four attributes: type, educational track, enrollment, and territorial criticality.
\begin{itemize}
    \item \textbf{School type.} The school type is determined jointly by the municipality type and a global proportion parameter 
$\rho _{CI} \in (0,1)$, set to $\rho _{CI} = 0.6$ across all experiments. Schools assigned to small municipalities are exclusively classified as CIs. For all other municipalities, the type is drawn as CI with probability $\rho _{CI}$ and as USI with probability $1 - \rho _{CI}$.

\item \textbf{Educational track.} For USI schools, the educational track is drawn from a fixed categorical distribution: 
academic programs with probability $0.4$, technical programs with probability $0.3$, and vocational programs with probability $0.3$. CI schools are assigned a non-applicable track label, consistently with the absence of curricular differentiation at the pre-primary and primary education levels.

\item \textbf{Enrollment.} The number of enrolled students $s_i$ is drawn independently and uniformly from the integer interval $[300, 800]$, capturing the typical enrollment range observed in Italian school institutions across different territorial contexts.

\item \textbf{Territorial criticality.} The criticality level $g_i$ of school $i$ is inherited directly from the municipality to which the school is assigned.
\end{itemize}

\subsubsection{Optimization Parameter Settings}
\label{subsubsec:parameter_settings}

The experiments were performed by evaluating multiple combinations of cost coefficients and threshold parameters in order to analyze the sensitivity of the model under different policy scenarios.

The cost coefficients were varied over predefined values to capture different relative priorities among the policy objectives. Specifically, all cost coefficients $C_1,\ldots,C_6$ were tested at $20$ and $80$, representing low and high penalty levels, respectively. The coefficients $C_1$ and $C_2$ account for inter-municipality aggregation and educational track heterogeneity, respectively, while $C_3$ and $C_4$ penalize aggregations involving schools in areas with territorial criticality levels 3 and 4. Finally, $C_5$ and $C_6$ represent the additional policy-related cost components.

The enrollment threshold coefficient was tested over two values, that is $
\gamma \in \{0.925, 0.95\}$,
to evaluate different levels of strictness in the minimum enrollment requirement.

The associated parameters were set to $\eta_{CI} \in \{0.3,0.4\}$, $ \eta_{USI} \in \{0.6,0.7\}$.
Lower values of $\eta$ generate stricter travel-time limits, while higher values allow more flexible aggregation opportunities. The selected ranges were chosen to reflect realistic policy trade-offs between student accessibility (travel time) and institutional consolidation (optimizing resource allocation and reducing the number of redundant autonomous school units).
The complete set of parameters used for the generation of the synthetic benchmark instances and the computational experiments is reported in \ref{app:instance_params} (Table \ref{tab:instance_params}). To provide a visual representation of the generated synthetic territories, the spatial distribution of schools across the four considered instance sizes ($n \in \{250, 500, 750, 1000\}$) is reported in \ref{app:instances_visualization}. As illustrated in Figure~\ref{fig:comparison_instances}, the bounding box expands proportionally with $n$, successfully maintaining a stable average spatial density while increasing the overall combinatorial complexity of the optimization landscape.

\subsection{Numerical Results}
\label{sec:numerical_results}

This section presents the computational results of the proposed optimization framework. The analysis is divided into two parts. First, the performance of the classical approach is evaluated and used as an exact benchmark. Second, the quantum and hybrid quantum-classical approaches are analyzed and compared with the classical benchmark to assess their current capabilities and limitations when applied to the school aggregation problem.

\subsubsection{Classical Optimization}
The computational results obtained with the classical optimization approach are presented in this section. 
The objective is to evaluate the computational behaviour of the formulation across increasing benchmark sizes and to investigate the impact of the policy parameters on the resulting school aggregation plans.

The experimental campaign considers four benchmark sizes ($n\in\{250,500,750,1000\}$) and more than one thousand optimization runs obtained by combining different values of the objective-function coefficients, accessibility thresholds, and dimensioning parameters. For every tested configuration, Gurobi finds an optimal solution, 
providing evidence that the formulation can be solved efficiently for the generated benchmark instances and policy configurations.

In the following, the computational behaviour is analyzed separately for each benchmark size, while the final subsection discusses scalability and model consistency across the experimental campaign. The analysis is divided into two stages: the first examines solver performance, solution characteristics, and parameter effects for each benchmark size; the second compares the results across sizes, focusing on model scalability and the relative importance of the decision parameters.

The detailed computational results for each benchmark size are reported in Tables~\ref{tab:classical_results_250}, \ref{tab:classical_results_500}, \ref{tab:classical_results_750}, and \ref{tab:classical_results_1000} in Appendix~\ref{classical_table}, while Table~\ref{tab:classical_summary} provides a comparative summary of the results and supports the cross-size analysis.

\subsubsubsection{Results by instance size}
This part of the work presents the computational results for each benchmark size separately in order to evaluate how the proposed optimization model behaves as the size of the school network increases. For each instance size, the analysis focuses on three main aspects: (i) the computational performance of the exact solver, (ii) the characteristics of the optimal solutions, and (iii) the sensitivity of the objective function to the parameters, assessed through the Pearson correlation coefficients \citet{pearson1895note}. For each parameter, the correlation coefficient is computed between its values across all tested configurations and the corresponding optimal objective values. Specifically, let $x=(x_1,\ldots,x_m)$ denote the values assumed by a given parameter over the $m$ tested instances and let $z=(z_1,\ldots,z_m)$ be the corresponding optimal objective values. The Pearson correlation coefficient is computed as:
\begin{equation}
r=\frac{ \sum\limits_{i=1}^{m}(x_i-\bar{x})(z_i-\bar{z})}
{\sqrt{\sum\limits_{i=1}^{m}(x_i-\bar{x})^2}\sqrt{\sum\limits_{i=1}^{m}(z_i-\bar{z})^2}}.
\end{equation}
Positive values indicate that increasing the corresponding parameter tends to increase the objective value, whereas negative values indicate the opposite. Correlation coefficients close to zero suggest that the parameter has little influence on the optimal solution within the considered experimental settings.

\paragraph{Instance size n=250.}
The smallest benchmark instances provide an initial assessment of the behaviour of the proposed model under relatively limited combinatorial complexity. Across the 512 tested parameter configurations, Gurobi successfully identified an optimal solution, confirming both the correctness of the mathematical formulation and the numerical stability of the model. The computational effort is extremely limited. The average solution time is only 0.032 s, with a median equal to 0.020 s and a maximum observed runtime of 1.05 s. 
The optimal objective value exhibits a considerable variability, ranging from 520 to 3040 with an average value of 1161.25. This wide interval indicates that the optimization model is highly sensitive to the policy priorities imposed through the objective-function coefficients, while remaining computationally insensitive to these variations. 
The correlation analysis provides additional insights into the contribution of each policy parameter. The strongest positive correlation is observed for the inter-municipality penalty coefficients $C_1$ and $C_6$ (Pearson coefficient equal to 0.449), indicating that increasing these penalties systematically increases the overall objective value by discouraging geographically dispersed aggregation patterns. The second most influential parameters are the curricular compatibility coefficients $C_2$ and $C_5$, whose correlation with the objective value is approximately 0.317. Conversely, the territorial protection coefficient $C_3$ shows no statistical influence on the final objective value, suggesting that the synthetic instances contain relatively few alternative solutions in which this penalty becomes active.
Among all parameters, the only one exhibiting a negative relationship with the objective value is the consolidation threshold $\gamma$ (correlation equal to $-0.353$). Larger values of $\gamma$, corresponding to less aggressive reorganization policies, reduce the number of feasible aggregation decisions and consequently produce solutions with lower objective values. Finally, the accessibility parameters $\eta_{CI}$ and $\eta_{USI}$ present correlations numerically equal to zero, indicating that, for instances of this size, the travel-time thresholds do not significantly modify the feasible assignment graph generated by the benchmark instances.
\paragraph{Instance size n=500.}
Doubling the number of schools considerably enlarges the feasible search space and increases the number of binary decision variables and assignment constraints. Nevertheless, the computational performance remains remarkably stable. The average solution time increases only from 0.032 s to 0.053 s, while the median runtime remains equal to 0.050 s. The maximum observed execution time is only 0.51 s, which is even lower than the largest runtime recorded for the smallest benchmark set, highlighting the robustness of the branch-and-bound search.
The average optimal objective value increases almost proportionally with the instance size, reaching 2290.63, with values ranging between 1000 and 6080. This behaviour confirms that the objective function scales consistently with the number of schools involved in the aggregation process.
The sensitivity analysis reveals a remarkably stable parameter hierarchy. The inter-municipality penalties ($C_1$ and $C_6$) remain the dominant drivers of the optimization process, exhibiting the largest correlation coefficients (0.445). Their influence is substantially stronger than that of all the remaining policy coefficients, demonstrating that geographical coherence represents the primary determinant of the final aggregation strategy.

The curricular compatibility coefficients ($C_2$ and $C_5$) continue to show a moderate influence (correlation approximately equal to 0.308), whereas the territorial fragility coefficient remains statistically negligible. Interestingly, the influence of the consolidation parameter $\gamma$ becomes slightly stronger than for the smaller instances (correlation equal to $-0.382$), suggesting that the proportion of schools allowed to remain autonomous plays an increasingly relevant role as the solution space expands. Also in this case, the accessibility thresholds exhibit practically zero correlation with the objective value, indicating that the travel-time constraints are rarely binding under the adopted synthetic territorial configuration.
\paragraph{Instance size n=750.}
Despite the substantial increase in combinatorial complexity, Gurobi continues to solve every tested instance to proven optimality, demonstrating the scalability of the proposed formulation.
The average computational time increases to only 0.086 s, with a median runtime of 0.070 s. Even for the most difficult solved instance, Gurobi requires only 1.65 s to certify optimality. Considering the size of the binary optimization model, these runtimes confirm that the compact formulation effectively limits the exploration of infeasible branches during the branch-and-bound process.
The optimal objective values range from 1520 to 9120, with an average value of 3461.21 and a standard deviation equal to 1887.33. The approximately linear growth of the objective function with respect to the number of schools confirms that the proposed cost structure scales consistently across different territorial dimensions.
The correlation analysis remains highly consistent with the previous benchmark sizes. The strongest influence is again associated with the geographical aggregation penalties ($C_1$ and $C_6$), whose correlation reaches approximately 0.452, making them the dominant parameters throughout the entire computational campaign. The curricular compatibility coefficients maintain a moderate influence (approximately 0.306), while the territorial protection coefficients remain statistically negligible.
The consolidation threshold $\gamma$ continues to represent the only parameter negatively correlated with the objective value (correlation equal to $-0.373$), confirming that stricter requirements on school autonomy naturally reduce the number of aggregation opportunities and consequently decrease the overall optimization cost.
Finally, the accessibility parameters again exhibit correlations extremely close to zero, suggesting that, within the generated benchmark instances, the feasible travel-time intervals are sufficiently wide and therefore do not significantly restrict the optimization process.

\paragraph{Instance size n=1000.}

The benchmark with 1000 schools represents the largest and most computationally demanding instance considered in this study. Despite the substantial increase in the size of the optimization model, Gurobi consistently identifies an optimal solution for every tested parameter configuration, confirming the robustness and scalability of the proposed formulation.

The formulation contains between 4\,923 and 6\,421 binary decision variables, depending on the accessibility thresholds adopted, while the number of constraints remains equal to 1\,129 for all tested scenarios. Nevertheless, the computational effort remains remarkably limited. The average solution time is approximately 0.11~s, with only a few isolated configurations requiring longer execution times. These results indicate that enlarging the school network does not produce a proportional increase in computational complexity, demonstrating that the branch-and-bound procedure continues to efficiently exploit the sparsity of the feasible assignment graph.

The sensitivity analysis confirms the trends already observed for the smaller benchmark sizes. The geographical aggregation coefficients ($C_1$ and $C_6$) remain the dominant drivers of the optimization process, exhibiting Pearson correlation coefficients of approximately 0.46 with the optimal objective value. This result confirms that the optimization primarily balances geographical coherence when constructing the aggregation plan. The curricular compatibility coefficients ($C_2$ and $C_5$) maintain a moderate positive influence (approximately 0.30), while the enrollment threshold parameter $\gamma$ remains the only parameter showing a systematic negative correlation ($r\approx-0.37$), indicating that stricter autonomy requirements naturally reduce the attainable objective value by limiting the number of feasible aggregations.

Conversely, the territorial protection coefficient ($C_3$), the administrative coefficient ($C_4$), and the accessibility thresholds ($\eta_{IC}$ and $\eta_{IS}$) exhibit correlation coefficients close to zero. Under the adopted benchmark generation procedure, these parameters rarely become binding and therefore have only a marginal influence on the optimal objective value. Overall, the results for $n=1000$ further confirm that the proposed formulation preserves both its computational efficiency and the stability of the policy parameter hierarchy even when applied to substantially larger school networks.

\subsubsubsection{Cross-Size Analysis}

The analysis presented in the previous subsections demonstrates that the proposed formulation exhibits a highly consistent behaviour across all benchmark sizes. Although the complexity of the school aggregation problem increases substantially as the number of schools grows, the computational performance, the structure of the optimal solutions, and the relative influence of the parameters remain remarkably stable. This subsection compares the four benchmark sets ($n=250$, $n=500$, $n=750$, and $n=1000$) in order to evaluate the scalability of the proposed formulation and to identify the general trends that characterize the optimization process. 

The comparison across the four benchmark sizes highlights the robustness of the proposed optimization framework from both a computational and a managerial perspective.

From a computational standpoint, the formulation exhibits excellent scalability. As expected, increasing the number of schools enlarges the combinatorial search space and substantially increases the number of binary decision variables and feasible assignment arcs. Nevertheless, the computational effort grows only moderately. The average solution time increases from 0.032~s for the smallest instances to 0.053~s, 0.086~s, and approximately 0.104~s for the largest benchmark. Even for the instances with 1000 schools, the average runtime remains close to 0.1~s, confirming that the branch-and-bound procedure efficiently exploits the sparsity of the optimization model. The institutional compatibility constraints, enrollment requirements, and accessibility restrictions considerably reduce the density of feasible assignment arcs, allowing the solver to prune large portions of the search space.

The objective function exhibits an almost perfectly linear scaling with the size of the school network. The average objective value increases from 1161.25 for $n=250$ to 2290.63, 3461.21, and 4568.13 for the larger benchmark sets. This behaviour indicates that the proposed cost structure behaves consistently across instances of different territorial size. Overall, the results suggest that increasing the size of the instance does not substantially change the behaviour of the proposed formulation.

The sensitivity analysis further confirms the stability of the optimization model. Across all benchmark sizes, the geographical aggregation coefficients ($C_1$ and $C_6$) consistently represent the dominant policy drivers, with Pearson correlation coefficients remaining remarkably stable around 0.45--0.46. This result indicates that geographical coherence systematically represents the primary factor influencing the optimal aggregation strategy. The enrollment threshold parameter $\gamma$ remains the only coefficient exhibiting a negative correlation with the objective value (approximately $r=-0.37$), confirming that stricter autonomy requirements reduce the number of feasible aggregations and consequently decrease the attainable objective value. The curricular compatibility coefficients ($C_2$ and $C_5$) maintain a moderate but highly stable influence, with correlation values close to 0.30 throughout the entire experimental campaign.

Conversely, the territorial protection coefficient ($C_3$), the administrative coefficient ($C_4$), and the accessibility thresholds ($\eta_{CI}$ and $\eta_{USI}$) exhibit correlation coefficients consistently close to zero. Within the generated benchmark instances, variations in these parameters have only a limited effect on the structure and cost of the optimal solutions. Their negligible statistical influence remains unchanged even for the largest problem dimension, suggesting that the optimization process is primarily governed by geographical compatibility and enrollment policies rather than by accessibility or territorial protection factors.

\subsubsection{Quantum Optimization}

This subsection evaluates the proposed formulation using the quantum optimization framework. The quantum computational campaign focuses on the benchmark instance with $n=500$ schools. This benchmark represents a realistic planning scenario for many Italian regional school systems and therefore provides a meaningful case study for assessing the proposed quantum optimization framework. The analysis is restricted to this instance due to limited access to the quantum solver and the computational resources available for hybrid quantum optimization.

The experimental design also adopts fixed values for the territorial protection coefficients ($C_3$ and $C_4$) and the accessibility thresholds ($\eta_{CI}=0.4$ and $\eta_{IS}=0.7$).
This choice is directly supported by the sensitivity analysis performed for the classical computation, where these parameters consistently exhibit Pearson correlation coefficients close to zero across all benchmark sizes. Their negligible statistical influence on the optimal objective value indicates that varying them does not significantly affect the optimization behaviour. Therefore, fixing these parameters allows the computational effort to be concentrated on the policy coefficients that effectively drive the optimization process without compromising the representativeness of the experimental analysis.

Table \ref{tab:quantum_results} in \ref{quantum_table} reports the results obtained with the hybrid quantum-classical optimization approach for the 32 experimental configurations considered in this study. Each configuration is executed five independent times in order to assess both the quality of the obtained solutions and the stability of the quantum optimization process. 


The experimental results demonstrate the robustness of the proposed quantum optimization framework. Across all 32 experimental instances, the hybrid solver consistently reproduces the optimal objective value obtained with classical computation. Consequently, the average objective value obtained over the five independent executions exactly matches the classical optimum in every configuration, while the corresponding standard deviation remains equal to zero. 
This result demonstrates that, for the considered benchmark instances, the D-Wave hybrid solver is able to recover the known global optimum. 

The average quantum execution time ranges approximately between 11.4~s and 12.8~s, while the corresponding standard deviation generally remains below one second. 


The results show that, despite the limited access to current quantum technology and the non-trivial size of the considered instances, the proposed formulation is well suited to hybrid quantum optimization. In particular, the hybrid quantum solver consistently identifies the same optimal objective value obtained by Gurobi across all tested instances, demonstrating that a carefully structured formulation can enable current quantum optimization technology to effectively solve instances of meaningful size.

\section{Real-World Validation: the Calabria School Network}\label{real}

The Calabria Region represents a particularly relevant case study for the school dimensioning problem because it combines demographic decline, territorial fragmentation, and significant accessibility constraints. Located in Southern Italy, Calabria is characterized by a predominantly mountainous morphology and a settlement structure composed of numerous small municipalities dispersed over a wide geographical area. These characteristics make the provision of public services, including education, particularly challenging, as maintaining a balance between administrative efficiency and territorial accessibility becomes increasingly difficult.

The regional school network is therefore affected by multiple and often conflicting planning objectives. On the one hand, the progressive reduction in the school-age population and the regulatory requirements on minimum enrolment encourage the aggregation of school institutions. On the other hand, the presence of inner and peripheral areas, together with the limited accessibility of several municipalities, requires preserving an adequate territorial coverage of educational services. These characteristics make Calabria an appropriate real-world benchmark for evaluating optimization models aimed at supporting school network reorganization, as the resulting decisions must simultaneously account for institutional constraints, geographical accessibility, and territorial equity.

This part of the work is organized into two parts. The first part describes the real-world optimization instance derived from the regional school network, including the dataset adopted and the information used to construct the optimization model. The second part presents the computational results obtained by applying both the classical and the hybrid quantum optimization approaches to the Calabria case study, allowing the assessment of the proposed methodology in a realistic operational planning scenario.

\subsection{Description of the instances}

The real-world case study considers the complete public school network of the Calabria Region. The dataset includes all state school institutions operating during the 2025/2026 school year 
and covers the five provincial areas of Catanzaro, Cosenza, Crotone, Reggio Calabria, and Vibo Valentia. For each institution, the available information includes the school identifier, the municipality, the school type, and the total number of enrolled students. The complete dataset is reported in \ref{app:schools}, specifically in Tables~\ref{tab:cz}-\ref{tab:cs}, where schools are grouped according to their province. 

The resulting dataset provides a comprehensive representation of the regional educational system and constitutes the basis for constructing the optimization instance analyzed in the following subsection.

\subsection{Computational Results}

We present the computational results obtained by applying the proposed optimization framework to the Calabria school network. The analysis aims to evaluate the effectiveness of the model in a real operational context and to verify whether the computational behaviour observed for the synthetic benchmark instances is confirmed when considering an actual regional school system.

To provide a comprehensive assessment, the results are discussed separately for the classical and the hybrid quantum optimization approaches. The first part analyzes the solutions obtained by solving the formulation with Gurobi, focusing on the characteristics of an optimal school aggregation plan and the corresponding computational performance. The second part evaluates the hybrid quantum implementation in order to assess the applicability of the proposed quantum optimization framework to a real-world decision-making problem.

\subsubsection{Classical Optimization}

The proposed formulation is solved by considering the complete Calabria school network under 64 different policy configurations obtained by varying the objective-function coefficients and the autonomy threshold. The detailed computational results for all 64 policy configurations are reported in \ref{app:calabria_results} in Table~\ref{tab:calabria_classical_full}, while a summary of the main structural indicators obtained across all configurations is presented in Table~\ref{tab:solution_analysis}.

The solver successfully identifies an optimal solution for every tested configuration, confirming the robustness of the proposed formulation when applied to a real-world planning problem. The instance remains unchanged throughout the experimental campaign and consists of 190 binary decision variables and 203 constraints, while only the policy parameters are modified. Consequently, the computational effort remains extremely limited, with solution times always below 0.20~s and generally around 0.05~s, demonstrating that the formulation can efficiently support real regional planning processes.

The objective function varies between 560 and 3760, reflecting the different policy priorities associated with the weighting coefficients. Consistently with the synthetic benchmark analysis, increasing the geographical aggregation penalties produces higher objective values, whereas increasing the autonomy threshold ($\gamma=0.95$) systematically reduces the objective function by restricting the number of feasible aggregation opportunities.

The sensitivity analysis confirms the ranking of the policy parameters previously observed on the synthetic instances. The geographical coefficients remain the principal drivers of the optimization process, with Pearson correlation coefficients equal to 0.57 for $C_6$ and 0.35 for $C_1$. The territorial coefficient $C_3$ exhibits a moderate positive influence ($r=0.32$), while the curricular compatibility coefficient $C_2$ has a comparatively smaller effect ($r=0.15$). As expected, the autonomy threshold $\gamma$ is the only parameter negatively correlated with the objective value ($r=-0.60$), confirming that more restrictive enrollment requirements reduce the number of admissible aggregations.

A noteworthy aspect of the Calabria case study concerns the exclusion of the coefficient $C_4$ from the computational analysis. The real regional dataset does not include school institutions classified as being at risk of dimensioning according to the adopted regulatory framework. Consequently, the corresponding penalty term is never activated in the objective function, making the inclusion of $C_4$ unnecessary for the analysis. This result further demonstrates that the proposed optimization framework naturally adapts to the characteristics of the analyzed school network, considering only those policy components that are effectively relevant to the specific territorial context.

We analyze the characteristics of the obtained solutions in terms of territorial distribution, track compatibility, and assignment criticality
\begin{itemize}
    \item \textbf{Cross-municipality aggregations.} The geographical distribution of cross-municipality aggregations remains remarkably stable across all 64 policy configurations, indicating that the territorial organization of the regional school network is primarily driven by its intrinsic geographical characteristics rather than by the adopted policy weights.
Among the five provinces, Cosenza consistently records the largest number of inter-municipality aggregations, varying between 2 and 9, with most scenarios generating between 6 and 8 aggregations. This behaviour reflects the large territorial extension of the province together with the high fragmentation of its municipal network. Catanzaro exhibits between 2 and 5 cross-municipality aggregations, whereas Crotone varies from 0 to 4 and Reggio Calabria from 2 to 3. Conversely, no inter-municipality aggregation is ever generated in the Province of Vibo Valentia, independently of the selected policy configuration. 
Increasing the geographical coefficients ($C_1$ and $C_6$) generally produces a moderate increase in the number of inter-municipality aggregations, although the overall territorial pattern remains substantially unchanged. This result confirms that the optimization process modifies only marginally the administrative organization while preserving the natural geographical structure of the regional school network.

\item \textbf{Educational track compatibility.} The optimization framework systematically favours aggregations involving schools with compatible educational tracks. Across all experiments, Cosenza records between 2 and 7 compatible aggregations, Catanzaro between 0 and 3, Crotone between 0 and 4, and Reggio Calabria between 0 and 2, whereas no compatible aggregation is required in Vibo Valentia. The number of incompatible aggregations always remains limited. The maximum observed values are equal to five in Cosenza, four in Catanzaro and Crotone, and three in Reggio Calabria. In several scenarios, particularly when the curricular compatibility coefficient assumes its highest value, incompatible aggregations completely disappear in one or more provinces. These results indicate that the compatibility component of the objective function effectively drives the optimization process towards pedagogically coherent school organizations while preserving the feasibility of the regional aggregation plan.

\item \textbf{Educational hubs.} The number of educational hubs generated by the optimization model exhibits only limited variability across the tested scenarios. Cosenza consistently contains the largest number of hubs, ranging from 5 to 10, while Catanzaro varies between 3 and 6 hubs. Crotone generates between 0 and 4 hubs and Reggio Calabria between 2 and 4. As observed for the cross-municipality aggregations, no educational hub is created in Vibo Valentia. The modest variation observed across the different policy configurations suggests that modifying the weighting coefficients mainly changes the allocation of individual school aggregations rather than the overall territorial organization of the educational network. Consequently, significant variations in the objective-function coefficients do not substantially change the structure of the solutions generated by the proposed formulation.

\item \textbf{Territorial criticality.} The analysis of the territorial criticality indicators confirms the robustness of the obtained solutions. The average provincial criticality remains relatively stable throughout the experimental campaign. Catanzaro varies between 0.67 and 2.00, Crotone between 0 and 2.75, Reggio Calabria between 1.00 and 1.50, and Cosenza between 0.75 and 2.00. In contrast, the Province of Vibo Valentia constantly reports a value equal to zero, confirming the absence of structurally critical schools requiring additional interventions. 
The regional distribution of municipalities among the five criticality classes also remains remarkably stable. The number of municipalities belonging to Class~0 varies between 3 and 7, whereas Class~1 ranges from 1 to 3 municipalities. Class~2 consistently includes between 4 and 6 municipalities, while Class~3 varies between 0 and 10 across the different policy configurations. Finally, no municipality is ever classified in Class~4 throughout the entire experimental campaign. 
These results indicate that the optimization model does not alter the underlying territorial criticality profile, which is determined by exogenous geographical and socio-demographic factors, but rather identifies aggregation plans that are consistent with the existing spatial distribution of territorial criticalities. Consequently, the generated aggregation plans do not introduce additional territorial imbalances in the distribution of educational services while highlighting those geographical areas where complementary policy measures may be required.

\end{itemize}

The limited variability observed in the geographical organization, the educational compatibility indicators, the number of hubs, and the territorial criticality demonstrates that the obtained aggregation plans are structurally robust with respect to changes in the policy parameters. 
From a decision-making perspective, this robustness is particularly relevant. Regional authorities may modify the relative importance assigned to geographical proximity, curricular compatibility, or territorial protection without substantially altering the overall organization of the school network. Consequently, the proposed optimization model supports policy makers not only in identifying globally optimal solutions but also in evaluating the stability and practical implications of alternative school dimensioning strategies.

\subsubsection{Quantum Optimization}

The proposed formulation was also solved using the hybrid quantum--classical framework for the complete Calabria school network under the same 64 policy configurations considered in the classical optimization experiments. The detailed computational results are reported in \ref{quantum_table_case} (Table~\ref{tab:calabria_quantum_full}).

Across all tested configurations, the solver consistently reproduced the optimal solutions obtained with classical computation. In every experiment, the average objective value obtained over the repeated quantum executions exactly coincides with the optimal solution, while the corresponding standard deviation remains equal to zero. Consequently, the optimality gap is always equal to 0\%, confirming that the hybrid quantum solver is able to recover the optimal aggregation plan for every policy configuration considered in the Calabria case study.

\section{Conclusions}\label{sec:conclusions} 
This paper presented a novel optimization framework for school network reorganization that jointly integrates institutional regulations, territorial accessibility, educational compatibility, and administrative constraints within a unified ILP formulation. Unlike traditional school planning approaches, the proposed model explicitly incorporates the regulatory requirements governing school dimensioning while preserving the geographical and educational characteristics of the resulting aggregation plans.

The computational study demonstrated the effectiveness of the proposed methodology on both synthetic benchmark instances and a real-world case study based on the complete school network of the Calabria Region. The classical solver consistently identified optimal solutions within negligible computational times, even for large-scale instances, confirming the computational efficiency and scalability of the proposed formulation. The extensive sensitivity analysis further highlighted the relative importance of the different policy parameters, showing that geographical coherence and enrollment thresholds represent the primary drivers of the aggregation process, whereas the resulting school networks remain structurally robust under different policy scenarios.

A second contribution of this work concerns the investigation of hybrid quantum optimization for school network planning. The experimental results demonstrated that the hybrid quantum framework is capable of consistently reproducing the optimal solutions obtained by classical computation. This confirms the suitability of the school dimensioning problem for emerging quantum optimization paradigms.

Beyond its computational contribution, the proposed framework provides a practical decision-support tool for regional educational authorities. By simultaneously balancing administrative efficiency, territorial equity, and educational coherence, the model enables policy makers to evaluate alternative school aggregation strategies while preserving the long-term sustainability of educational services, particularly in geographically fragile territories affected by demographic decline.

Several research directions naturally emerge from this work. From a methodological perspective, future developments could incorporate additional planning dimensions, including multi-period decision horizons, demographic forecasting, budget constraints, and uncertainty in future student enrollment. The optimization framework could also be extended to explicitly consider transportation networks, environmental impacts, and multi-objective formulations capable of simultaneously optimizing efficiency, accessibility, and equity criteria. 
From the computational perspective, future research will investigate larger-scale quantum implementations as next-generation quantum hardware becomes available. In particular, improvements in qubit connectivity, problem embedding techniques, and hybrid quantum-classical algorithms may substantially enhance the applicability of quantum optimization to large public-service planning problems. The proposed school dimensioning model therefore represents a promising benchmark for evaluating the evolution of quantum optimization technologies in complex real-world decision-support applications.

\section*{Declaration of competing interest}
The authors declare that they have no known competing financial interests or personal relationships that could have appeared to
influence the work reported in this paper.

\section*{Data availability}
Data will be made available on request.

  \bibliography{bibliography}

\newpage

\appendix

\section{Instances Visualization}
\label{app:instances_visualization}

\begin{figure}[htbp]
    \centering
    
    \caption{Spatial distribution of the synthetic territorial instances for $n = 250, 500, 750, 1000$. Each point represents a school node plotted within its respective $[0, n] \times [0, n]$ bounding box. Markers are differentiated by shape and color to reflect the underlying hierarchical settlement attributes: large markers (squares and diamonds) located in wider clusters correspond to provincial capitals and large urban hubs, whereas high-density, tight clusters of smaller circular and triangular markers represent intermediate and peripheral municipalities. Color shading transitions from cooler tones (blues/greys) to warmer tones (yellows/reds) to visually map the progressive levels of territorial and socioeconomic criticality ($g_i \in \{0, 1, 2, 3, 4\}$).}
    \label{fig:comparison_instances}
    \includegraphics[width=0.85\textwidth]{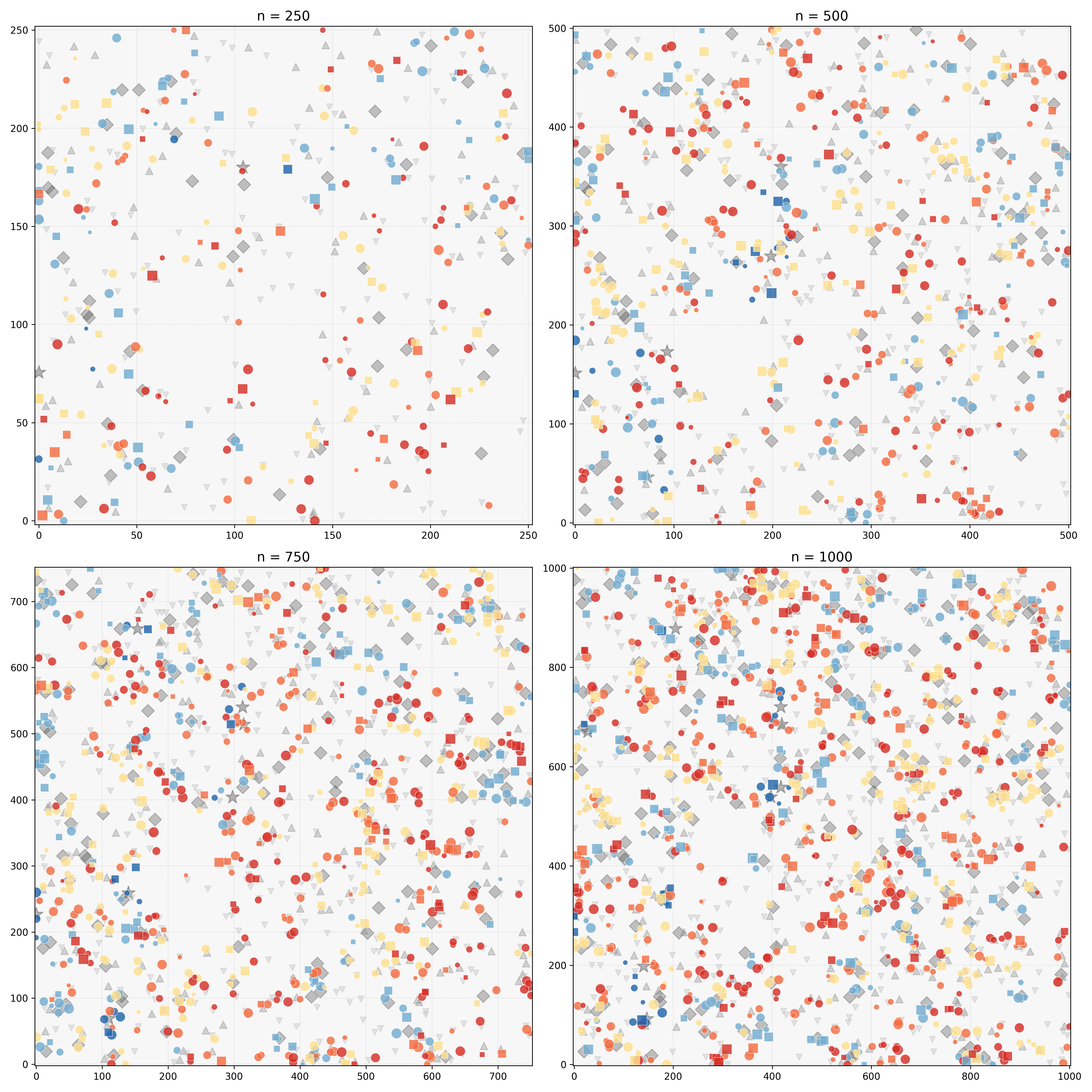}
\end{figure}

This appendix provides a graphical insights into the synthetic benchmarks generated for the computational study. Figure~\ref{fig:comparison_instances} displays the geometric properties of the generated spatial configurations.

\clearpage
\section{Synthetic Instance Generation Parameters}\label{app:instance_params} 

\begin{table}[ht]
\centering
\caption{Structural parameters of the four municipality types in the synthetic instance generator. Allocation weights govern the probability of a school being assigned to each municipality type and are normalized to sum to one across all municipalities of each type.}
\label{tab:municipality_types}
\begin{tabular}{lcccc}
\toprule
\textbf{Municipality type} & \textbf{Cardinality} & \textbf{Radius} $r$ 
& \textbf{Criticality} $g$ & \textbf{Allocation weight} \\
\midrule
Provincial capital  & $\max(1,\lfloor 0.01n \rfloor)$ & 20 & 0         & 5 \\
Large municipality  & $\lfloor n/6 \rfloor$            & 15 & $\{1,2\}$ & 4 \\
Medium municipality & $\lfloor n/4 \rfloor$            & 10 & $\{2,3,4\}$ & 2 \\
Small municipality  & $\lfloor n/2 \rfloor$            & 5  & $\{3,4\}$ & 1 \\
\bottomrule
\end{tabular}
\end{table}

\begin{table}[h]
\centering
\small
\caption{Summary of parameters used in synthetic instance generation and optimization experiments.}
\label{tab:instance_params}
\begin{tabular}{p{6cm} p{9cm}}
\toprule
\textbf{Parameter} & \textbf{Value / Formula} \\
\midrule

Instance size & $n \in \{250,500,750,1000\}$ \\

Spatial domain & $[0,n]\times[0,n]$ \\

Number of capitals & $\max(1,\lfloor 0.01n \rfloor)$ \\

Number of large municipalities & $\lfloor n/6 \rfloor$ \\

Number of medium municipalities & $\lfloor n/4 \rfloor$ \\

Number of small municipalities & $\lfloor n/2 \rfloor$ \\

Municipality centers & $(\bar{lat}_m,\bar{lon}_m)\sim U([0,n]\times[0,n])$ \\

School coordinates & 
$\mathcal{N}((\bar{lat}_m,\bar{lon}_m),r_m^2 I)$ \\

Allocation weights & Capitals: 5, Large: 4, Medium: 2, Small: 1 \\

CI proportion & $\rho_{CI}=0.6$ \\

School type & Small municipalities $\rightarrow$ CI only; otherwise Bernoulli($0.6$) \\

USI track distribution & Academic: 0.4, Technical: 0.3, Vocational: 0.3 \\

Enrollment & $s_i \sim U([300,800])$ \\

Criticality & $g \in \{0,\dots,4\}$ \\

Average speed & $v=1$ \\

Travel-time thresholds &
$T_{\max}^{CI},\; T_{\max}^{USI}$ \\

$\eta_{CI}$ & $\{0.3,0.4\}$ \\

$\eta_{USI}$ & $\{0.6,0.7\}$ \\

$\gamma$ & $\{0.925,0.95\}$ \\
$C_1$ & $\{20,80\}$ \\

$C_2$ & $\{20,80\}$ \\

$C_3$ & $\{20,80\}$ \\

$C_4$ & $\{20,80\}$ \\

$C_5$ & $\{20,80\}$ \\

$C_6$ & $\{20,80\}$ \\

\bottomrule
\end{tabular}

\end{table}

\clearpage
\section{Computational Results - Classical optimization}
\label{classical_table}
The following section presents the computational results obtained with the classical optimization model on the benchmark instances. The analysis is structured according to the four considered problem sizes, namely $n=250$, $n=500$, $n=750$, and $n=1000$ schools, in order to evaluate the computational behaviour of the model as the problem size increases. For each benchmark size, the results obtained under the different combinations of objective-function coefficients and policy parameters are first presented and discussed. Particular attention is then devoted to the computational time required to solve the instances, as well as to the effect of the main model parameters on the resulting objective values. Finally, the results across the four benchmark sizes are compared to assess the scalability and consistency of the classical optimization model and to identify how the computational performance evolves with increasing problem dimensions. The complete set of computational results is reported in the corresponding tables, while the discussion focuses on the main patterns emerging from the experimental analysis. These results also provide the classical reference solutions used in the subsequent evaluation of the hybrid quantum-classical approach.

\begin{table}[htbp]
\centering
\caption{Computational results obtained with the classical optimization model configurations of the $n=250$ benchmark instance.}
\label{tab:classical_results_250}
\resizebox{\textwidth}{!}{
\begin{tabular}{ccccccccccc||ccccccccccc||ccccccccccc||ccccccccccc}
\toprule
\cmidrule(lr){1-11}
\cmidrule(lr){12-22}
\cmidrule(lr){23-33}
\cmidrule(lr){34-44}
$C_1$ & $C_2$ & $C_3$ & $C_4$ & $C_5$ & $C_6$ & $\gamma$ & $\eta_{IC}$ & $\eta_{IS}$ & $z^C$ & \textbf{Time (s)} & $C_1$ & $C_2$ & $C_3$ & $C_4$ & $C_5$ & $C_6$ & $\gamma$ & $\eta_{IC}$ & $\eta_{IS}$ & $z^C$ & \textbf{Time (s)} & $C_1$ & $C_2$ & $C_3$ & $C_4$ & $C_5$ & $C_6$ & $\gamma$ & $\eta_{IC}$ & $\eta_{IS}$ & $z^C$ & \textbf{Time (s)} & $C_1$ & $C_2$ & $C_3$ & $C_4$ & $C_5$ & $C_6$ & $\gamma$ & $\eta_{IC}$ & $\eta_{IS}$ & $z^C$ & \textbf{Time (s)} \\
\midrule
20 & 20 & 20 & 20 & 20 & 20 & 0.925 & 0.3 & 0.6 & 760 & 0.02 & 80 & 20 & 80 & 20 & 20 & 80 & 0.95 & 0.3 & 0.7 & 1000 & 0.02 & 20 & 80 & 20 & 20 & 80 & 80 & 0.925 & 0.4 & 0.6 & 1900 & 0.02 & 80 & 80 & 80 & 80 & 20 & 20 & 0.95 & 0.4 & 0.6 & 520 & 0.03 \\
20 & 20 & 80 & 80 & 20 & 20 & 0.95 & 0.4 & 0.7 & 520 & 0.02 & 80 & 80 & 20 & 80 & 80 & 20 & 0.925 & 0.4 & 0.7 & 1900 & 0.03 & 80 & 20 & 20 & 20 & 20 & 20 & 0.925 & 0.3 & 0.7 & 760 & 0.03 & 20 & 20 & 80 & 20 & 20 & 80 & 0.95 & 0.3 & 0.7 & 520 & 0.02 \\
20 & 80 & 80 & 20 & 20 & 80 & 0.95 & 0.4 & 0.6 & 1300 & 0.02 & 20 & 20 & 20 & 20 & 80 & 80 & 0.925 & 0.4 & 0.6 & 760 & 0.02 & 80 & 20 & 80 & 80 & 20 & 20 & 0.95 & 0.4 & 0.7 & 520 & 0.02 & 20 & 80 & 20 & 80 & 80 & 20 & 0.95 & 0.3 & 0.6 & 520 & 0.02 \\
80 & 20 & 20 & 80 & 80 & 20 & 0.95 & 0.3 & 0.6 & 1300 & 0.02 & 20 & 80 & 20 & 20 & 20 & 20 & 0.925 & 0.3 & 0.7 & 760 & 0.03 & 80 & 80 & 80 & 20 & 20 & 80 & 0.95 & 0.3 & 0.7 & 1300 & 0.02 & 80 & 20 & 20 & 20 & 80 & 80 & 0.925 & 0.4 & 0.7 & 1900 & 0.02 \\
80 & 80 & 20 & 20 & 80 & 80 & 0.925 & 0.4 & 0.6 & 3040 & 0.03 & 20 & 80 & 80 & 80 & 20 & 80 & 0.925 & 0.3 & 0.6 & 1900 & 0.02 & 20 & 20 & 20 & 80 & 80 & 20 & 0.95 & 0.3 & 0.6 & 520 & 0.02 & 80 & 80 & 20 & 20 & 20 & 20 & 0.925 & 0.3 & 0.7 & 760 & 0.03 \\
\bottomrule
\end{tabular}
}
\end{table}

\begin{table}[htbp]
\centering
\caption{Computational results obtained with the classical optimization model configurations of the $n=500$ benchmark instance.}
\label{tab:classical_results_500}
\resizebox{\textwidth}{!}{
\begin{tabular}{ccccccccccc||ccccccccccc||ccccccccccc||ccccccccccc}
\toprule
\cmidrule(lr){1-11}
\cmidrule(lr){12-22}
\cmidrule(lr){23-33}
\cmidrule(lr){34-44}
$C_1$ & $C_2$ & $C_3$ & $C_4$ & $C_5$ & $C_6$ & $\gamma$ & $\eta_{IC}$ & $\eta_{IS}$ & $z^C$ & \textbf{Time (s)} & $C_1$ & $C_2$ & $C_3$ & $C_4$ & $C_5$ & $C_6$ & $\gamma$ & $\eta_{IC}$ & $\eta_{IS}$ & $z^C$ & \textbf{Time (s)} & $C_1$ & $C_2$ & $C_3$ & $C_4$ & $C_5$ & $C_6$ & $\gamma$ & $\eta_{IC}$ & $\eta_{IS}$ & $z^C$ & \textbf{Time (s)} & $C_1$ & $C_2$ & $C_3$ & $C_4$ & $C_5$ & $C_6$ & $\gamma$ & $\eta_{IC}$ & $\eta_{IS}$ & $z^C$ & \textbf{Time (s)} \\
\midrule
20 & 20 & 20 & 20 & 20 & 20 & 0.925 & 0.3 & 0.6 & 1520 & 0.03 & 20 & 80 & 20 & 20 & 20 & 20 & 0.925 & 0.3 & 0.6 & 1520 & 0.03 & 80 & 20 & 20 & 20 & 20 & 20 & 0.925 & 0.3 & 0.6 & 1520 & 0.05 & 80 & 80 & 20 & 20 & 20 & 20 & 0.925 & 0.3 & 0.6 & 1520 & 0.05 \\
20 & 20 & 20 & 20 & 20 & 20 & 0.925 & 0.3 & 0.7 & 1520 & 0.04 & 20 & 80 & 20 & 20 & 20 & 20 & 0.925 & 0.3 & 0.7 & 1520 & 0.04 & 80 & 20 & 20 & 20 & 20 & 20 & 0.925 & 0.3 & 0.7 & 1520 & 0.05 & 80 & 80 & 20 & 20 & 20 & 20 & 0.925 & 0.3 & 0.7 & 1520 & 0.05 \\
20 & 20 & 20 & 20 & 20 & 20 & 0.925 & 0.4 & 0.6 & 1520 & 0.06 & 20 & 80 & 20 & 20 & 20 & 20 & 0.925 & 0.4 & 0.6 & 1520 & 0.06 & 80 & 20 & 20 & 20 & 20 & 20 & 0.925 & 0.4 & 0.6 & 1520 & 0.06 & 80 & 80 & 20 & 20 & 20 & 20 & 0.925 & 0.4 & 0.6 & 1520 & 0.05 \\
20 & 20 & 20 & 20 & 20 & 20 & 0.925 & 0.4 & 0.7 & 1520 & 0.06 & 20 & 80 & 20 & 20 & 20 & 20 & 0.925 & 0.4 & 0.7 & 1520 & 0.05 & 80 & 20 & 20 & 20 & 20 & 20 & 0.925 & 0.4 & 0.7 & 1520 & 0.05 & 80 & 80 & 20 & 20 & 20 & 20 & 0.925 & 0.4 & 0.7 & 1520 & 0.06 \\
20 & 20 & 20 & 20 & 20 & 20 & 0.95 & 0.3 & 0.6 & 1000 & 0.04 & 20 & 80 & 20 & 20 & 20 & 20 & 0.95 & 0.3 & 0.6 & 1000 & 0.03 & 80 & 20 & 20 & 20 & 20 & 20 & 0.95 & 0.3 & 0.6 & 1000 & 0.04 & 80 & 80 & 20 & 20 & 20 & 20 & 0.95 & 0.3 & 0.6 & 1000 & 0.03 \\
\bottomrule
\end{tabular}
}
\end{table}

\begin{table}[htbp]
\centering
\caption{Computational results obtained with the classical optimization model for the $n=750$ benchmark instance.}
\label{tab:classical_results_750}
\resizebox{\textwidth}{!}{
\begin{tabular}{ccccccccccc||ccccccccccc||ccccccccccc||ccccccccccc}
\toprule
\cmidrule(lr){1-11}\cmidrule(lr){12-22}\cmidrule(lr){23-33}\cmidrule(lr){34-44}
$C_1$ & $C_2$ & $C_3$ & $C_4$ & $C_5$ & $C_6$ & $\gamma$ & $\eta_{IC}$ & $\eta_{IS}$ & $z^C$ & \textbf{Time (s)} & $C_1$ & $C_2$ & $C_3$ & $C_4$ & $C_5$ & $C_6$ & $\gamma$ & $\eta_{IC}$ & $\eta_{IS}$ & $z^C$ & \textbf{Time (s)} & $C_1$ & $C_2$ & $C_3$ & $C_4$ & $C_5$ & $C_6$ & $\gamma$ & $\eta_{IC}$ & $\eta_{IS}$ & $z^C$ & \textbf{Time (s)} & $C_1$ & $C_2$ & $C_3$ & $C_4$ & $C_5$ & $C_6$ & $\gamma$ & $\eta_{IC}$ & $\eta_{IS}$ & $z^C$ & \textbf{Time (s)} \\
\midrule
20 & 20 & 20 & 20 & 20 & 20 & 0.925 & 0.3 & 0.6 & 2280 & 0.07 & 20 & 80 & 20 & 20 & 20 & 20 & 0.925 & 0.3 & 0.6 & 2280 & 0.09 & 80 & 20 & 20 & 20 & 20 & 20 & 0.925 & 0.3 & 0.7 & 2280 & 0.04 & 80 & 80 & 20 & 20 & 20 & 20 & 0.925 & 0.3 & 0.7 & 2280 & 0.05 \\
20 & 20 & 20 & 20 & 20 & 20 & 0.925 & 0.3 & 0.7 & 2280 & 0.09 & 20 & 80 & 20 & 20 & 20 & 20 & 0.925 & 0.3 & 0.7 & 2280 & 0.08 & 80 & 20 & 20 & 20 & 20 & 20 & 0.925 & 0.4 & 0.6 & 2280 & 0.08 & 80 & 80 & 20 & 20 & 20 & 20 & 0.925 & 0.4 & 0.6 & 2280 & 0.06 \\
20 & 20 & 20 & 20 & 20 & 20 & 0.925 & 0.4 & 0.6 & 2280 & 0.09 & 20 & 80 & 20 & 20 & 20 & 20 & 0.925 & 0.4 & 0.6 & 2280 & 0.07 & 80 & 20 & 20 & 20 & 20 & 20 & 0.925 & 0.4 & 0.7 & 2280 & 0.08 & 80 & 80 & 20 & 20 & 20 & 20 & 0.925 & 0.4 & 0.7 & 2280 & 0.06 \\
20 & 20 & 20 & 20 & 20 & 20 & 0.925 & 0.4 & 0.7 & 2280 & 0.09 & 20 & 80 & 20 & 20 & 20 & 20 & 0.925 & 0.4 & 0.7 & 2280 & 0.16 & 80 & 20 & 20 & 20 & 20 & 20 & 0.95 & 0.3 & 0.6 & 1520 & 0.06 & 80 & 80 & 20 & 20 & 20 & 20 & 0.95 & 0.3 & 0.6 & 1520 & 0.06 \\
20 & 20 & 20 & 20 & 20 & 20 & 0.95 & 0.3 & 0.6 & 1520 & 0.07 & 20 & 80 & 20 & 20 & 20 & 20 & 0.95 & 0.3 & 0.6 & 1520 & 0.13 & 80 & 20 & 20 & 20 & 20 & 20 & 0.95 & 0.3 & 0.7 & 1520 & 0.09 & 80 & 80 & 20 & 20 & 20 & 20 & 0.95 & 0.3 & 0.7 & 1520 & 0.07 \\
\bottomrule
\end{tabular}
}
\end{table}

\begin{table}[htbp]
\centering
\caption{Computational results obtained with the classical optimization model for the benchmark instances with $n=1000$ schools. }
\label{tab:classical_results_1000}
\resizebox{\textwidth}{!}{
\begin{tabular}{ccccccccccc||ccccccccccc||ccccccccccc||ccccccccccc}
\toprule
\cmidrule(lr){1-11}\cmidrule(lr){12-22}\cmidrule(lr){23-33}\cmidrule(lr){34-44}
$C_1$ & $C_2$ & $C_3$ & $C_4$ & $C_5$ & $C_6$ & $\gamma$ & $\eta_{IC}$ & $\eta_{IS}$ & $z^C$ & \textbf{Time (s)} &
$C_1$ & $C_2$ & $C_3$ & $C_4$ & $C_5$ & $C_6$ & $\gamma$ & $\eta_{IC}$ & $\eta_{IS}$ & $z^C$ & \textbf{Time (s)} &
$C_1$ & $C_2$ & $C_3$ & $C_4$ & $C_5$ & $C_6$ & $\gamma$ & $\eta_{IC}$ & $\eta_{IS}$ & $z^C$ & \textbf{Time (s)} &
$C_1$ & $C_2$ & $C_3$ & $C_4$ & $C_5$ & $C_6$ & $\gamma$ & $\eta_{IC}$ & $\eta_{IS}$ & $z^C$ & \textbf{Time (s)} \\
\midrule
20 & 20 & 20 & 20 & 20 & 20 & 0.925 & 0.3 & 0.6 & 3000 & 0.05 & 20 & 80 & 20 & 20 & 20 & 20 & 0.925 & 0.3 & 0.6 & 3000 & 0.07 & 80 & 20 & 20 & 20 & 20 & 20 & 0.925 & 0.3 & 0.6 & 3000 & 0.05 & 80 & 80 & 20 & 20 & 20 & 20 & 0.925 & 0.3 & 0.6 & 3000 & 0.06 \\
20 & 20 & 20 & 20 & 20 & 20 & 0.925 & 0.3 & 0.7 & 3000 & 0.05 & 20 & 80 & 20 & 20 & 20 & 20 & 0.925 & 0.3 & 0.7 & 3000 & 0.08 & 80 & 20 & 20 & 20 & 20 & 20 & 0.925 & 0.3 & 0.7 & 3000 & 0.27 & 80 & 80 & 20 & 20 & 20 & 20 & 0.925 & 0.3 & 0.7 & 3000 & 0.05 \\
20 & 20 & 20 & 20 & 20 & 20 & 0.925 & 0.4 & 0.6 & 3000 & 0.06 & 20 & 80 & 20 & 20 & 20 & 20 & 0.925 & 0.4 & 0.6 & 3000 & 0.08 & 80 & 20 & 20 & 20 & 20 & 20 & 0.925 & 0.4 & 0.6 & 3000 & 0.16 & 80 & 80 & 20 & 20 & 20 & 20 & 0.925 & 0.4 & 0.6 & 3000 & 0.09 \\
20 & 20 & 20 & 20 & 20 & 20 & 0.925 & 0.4 & 0.7 & 3000 & 0.06 & 20 & 80 & 20 & 20 & 20 & 20 & 0.925 & 0.4 & 0.7 & 3000 & 0.09 & 80 & 20 & 20 & 20 & 20 & 20 & 0.925 & 0.4 & 0.7 & 3000 & 0.57 & 80 & 80 & 20 & 20 & 20 & 20 & 0.925 & 0.4 & 0.7 & 3000 & 0.12 \\
20 & 20 & 20 & 20 & 20 & 20 & 0.95 & 0.3 & 0.6 & 2000 & 0.05 & 20 & 80 & 20 & 20 & 20 & 20 & 0.95 & 0.3 & 0.6 & 2000 & 0.09 & 80 & 20 & 20 & 20 & 20 & 20 & 0.95 & 0.3 & 0.6 & 2000 & 0.09 & 80 & 80 & 20 & 20 & 20 & 20 & 0.95 & 0.3 & 0.6 & 2000 & 0.05 \\
\bottomrule
\end{tabular}
}
\end{table}

\begin{table}[htbp]
\centering
\caption{Summary statistics of the computational results obtained with the classical optimization model. For each benchmark size, the table reports the optimization model dimensions, the solver performance, the distribution of the optimal objective values, and the Pearson correlation coefficients used to quantify the influence of the parameters on the objective function across all 512 experimental scenarios.}
\label{tab:classical_summary}
\renewcommand{\arraystretch}{1.25}
\setlength{\tabcolsep}{6pt}
\begin{tabular}{lcccc}
\hline
\textbf{Performance Metric} & \textbf{$n=250$} & \textbf{$n=500$} & \textbf{$n=750$} & \textbf{$n=1000$} \\
\hline

\multicolumn{5}{l}{\textit{Optimization model}}\\
Binary variables & 1563--1985 & 2811--3819 & 3939--5287 & 4923--6421 \\
Constraints & 317 & 559 & 846 & 1129 \\

\hline
\multicolumn{5}{l}{\textit{Computational performance}}\\
Number of tested configurations & 512 & 512 & 512 & 512 \\
Average runtime (s) & 0.032 & 0.053 & 0.086 & 0.110 \\
Median runtime (s) & 0.020 & 0.050 & 0.070 & 0.090 \\
Maximum runtime (s) & 1.050 & 0.510 & 1.650 & 1.830 \\

\hline
\multicolumn{5}{l}{\textit{Objective function}}\\
Minimum objective value & 520 & 1000 & 1520 & 2040 \\
Maximum objective value & 3040 & 6080 & 9120 & 12160 \\
Mean objective value & 1161.25 & 2290.63 & 3461.21 & 4630.52 \\

\hline
\multicolumn{5}{l}{\textit{Parameter sensitivity (Pearson correlation coefficient)}}\\
Geographical penalties ($C_1,C_6$) & 0.449 & 0.445 & 0.452 & 0.460 \\
Curricular penalties ($C_2,C_5$) & 0.317 & 0.308 & 0.306 & 0.300 \\
Autonomy threshold ($\gamma$) & $-0.353$ & $-0.382$ & $-0.373$ & $-0.370$ \\
Territorial penalties ($C_3,C_4$) & $\approx0$ & $\approx0$ & $\approx0$ & $\approx0$ \\
Accessibility thresholds ($\eta_{CI},\eta_{USI}$) & $\approx0$ & $\approx0$ & $\approx0$ & $\approx0$ \\

\hline
\end{tabular}
\end{table}

\newpage
\section{Computational Results - Quantum optimization}
\label{quantum_table}

\begin{table}[ht]
\centering
\caption{Computational results obtained with the hybrid quantum optimization framework for the benchmark instances with $n=500$ schools. For each parameter configuration, the table reports the parameter values, the optimal objective value obtained by the classical optimization model ($z^{C}$), the corresponding optimal objective value averaged over five independent quantum executions ($\overline{z}^{Q}$), the standard deviation of the quantum objective values ($\sigma_{z}^{Q}$), the average quantum solution time ($\overline{t}^{Q}$), its standard deviation ($\sigma_{t}^{Q}$), and the optimality gap with respect to the classical solution.}
\label{tab:quantum_results}
\smallskip
\resizebox{0.85\textwidth}{!}{
\begin{tabular}{ccccccccc|c|cccc|c}
\hline
\multicolumn{9}{c|}{\textbf{Parameters}} & \textbf{Classical} & \multicolumn{4}{c|}{\textbf{Quantum}} & \multirow{2}{*}{\textbf{Gap}} \\ 
\cmidrule(lr){1-9}
\cmidrule(lr){10-10}
\cmidrule(lr){11-14}
$C_1$ & $C_2$ & $C_3$ & $C_4$ & $C_5$ & $C_6$ & $\gamma$ & $\eta_{IC}$ & $\eta_{IS}$ & $z^{C}$ & $\overline{z}^{Q}$ & $\sigma_{z}^{Q}$ & $\overline{t}^{Q}$ & $\sigma_{t}^{Q}$ & \\
\hline

    20    & 20    & 80    & 80    & 20    & 20    & 0.925 & 0.4   & 0.7   & 1520  & 1520& 0.00  & 12.45 & 1.43  & 0\% \\
    20    & 20    & 80    & 80    & 20    & 20    & 0.95  & 0.4   & 0.7   & 1000  & 1000& 0.00  & 11.69 & 0.70  & 0\% \\
    20    & 20    & 80    & 80    & 20    & 80    & 0.925 & 0.4   & 0.7   & 1520  & 1520& 0.00  & 11.47 & 0.18  & 0\% \\
    20    & 20    & 80    & 80    & 20    & 80    & 0.95  & 0.4   & 0.7   & 1000  & 1000& 0.00  & 11.48 & 0.16  & 0\% \\
    20    & 20    & 80    & 80    & 80    & 20    & 0.925 & 0.4   & 0.7   & 1520  & 1520& 0.00  & 11.66 & 0.22  & 0\% \\
    20    & 20    & 80    & 80    & 80    & 20    & 0.95  & 0.4   & 0.7   & 1000  & 1000& 0.00  & 11.66 & 0.36  & 0\% \\
    20    & 20    & 80    & 80    & 80    & 80    & 0.925 & 0.4   & 0.7   & 1520  & 1520& 0.00  & 11.74 & 0.19  & 0\% \\
    20    & 20    & 80    & 80    & 80    & 80    & 0.95  & 0.4   & 0.7   & 1000  & 1000& 0.00  & 11.48 & 0.23  & 0\% \\
    20    & 80    & 80    & 80    & 20    & 20    & 0.925 & 0.4   & 0.7   & 1520  & 1520& 0.00  & 11.92 & 0.24  & 0\% \\
    20    & 80    & 80    & 80    & 20    & 20    & 0.95  & 0.4   & 0.7   & 1000  & 1000& 0.00  & 11.71 & 0.33  & 0\% \\
    20    & 80    & 80    & 80    & 20    & 80    & 0.925 & 0.4   & 0.7   & 3800  & 3800& 0.00  & 11.52 & 0.08  & 0\% \\
    20    & 80    & 80    & 80    & 20    & 80    & 0.95  & 0.4   & 0.7   & 2500  & 2500& 0.00  & 11.68 & 0.55  & 0\% \\
    20    & 80    & 80    & 80    & 80    & 20    & 0.925 & 0.4   & 0.7   & 1520  & 1520& 0.00  & 11.74 & 0.90  & 0\% \\
    20    & 80    & 80    & 80    & 80    & 20    & 0.95  & 0.4   & 0.7   & 1000  & 1000& 0.00  & 11.66 & 0.35  & 0\% \\
    20    & 80    & 80    & 80    & 80    & 80    & 0.925 & 0.4   & 0.7   & 3800  & 3800& 0.00  & 11.50 & 0.25  & 0\% \\
    20    & 80    & 80    & 80    & 80    & 80    & 0.95  & 0.4   & 0.7   & 2500  & 2500& 0.00  & 11.93 & 1.02  & 0\% \\
    80    & 20    & 80    & 80    & 20    & 20    & 0.925 & 0.4   & 0.7   & 1520  & 1520& 0.00  & 11.72 & 0.27  & 0\% \\
    80    & 20    & 80    & 80    & 20    & 20    & 0.95  & 0.4   & 0.7   & 1000  & 1000& 0.00  & 11.66 & 0.22  & 0\% \\
    80    & 20    & 80    & 80    & 20    & 80    & 0.925 & 0.4   & 0.7   & 3320  & 3320& 0.00  & 11.91 & 0.48  & 0\% \\
    80    & 20    & 80    & 80    & 20    & 80    & 0.95  & 0.4   & 0.7   & 2020  & 2020& 0.00  & 12.84 & 2.18  & 0\% \\
    80    & 20    & 80    & 80    & 80    & 20    & 0.925 & 0.4   & 0.7   & 3800  & 3800& 0.00  & 11.87 & 0.53  & 0\% \\
    80    & 20    & 80    & 80    & 80    & 20    & 0.95  & 0.4   & 0.7   & 2500  & 2500& 0.00  & 11.58 & 0.50  & 0\% \\
    80    & 20    & 80    & 80    & 80    & 80    & 0.925 & 0.4   & 0.7   & 3800  & 3800& 0.00  & 12.01 & 0.42  & 0\% \\
    80    & 20    & 80    & 80    & 80    & 80    & 0.95  & 0.4   & 0.7   & 2500  & 2500& 0.00  & 11.42 & 0.30  & 0\% \\
    80    & 80    & 80    & 80    & 20    & 20    & 0.925 & 0.4   & 0.7   & 1520  & 1520& 0.00  & 12.49 & 1.03  & 0\% \\
    80    & 80    & 80    & 80    & 20    & 20    & 0.95  & 0.4   & 0.7   & 1000  & 1000& 0.00  & 11.91 & 1.24  & 0\% \\
    80    & 80    & 80    & 80    & 20    & 80    & 0.925 & 0.4   & 0.7   & 3800  & 3800& 0.00  & 11.91 & 0.55  & 0\% \\
    80    & 80    & 80    & 80    & 20    & 80    & 0.95  & 0.4   & 0.7   & 2500  & 2500& 0.00  & 11.76 & 0.51  & 0\% \\
    80    & 80    & 80    & 80    & 80    & 20    & 0.925 & 0.4   & 0.7   & 3800  & 3800& 0.00  & 12.21 & 0.67  & 0\% \\
    80    & 80    & 80    & 80    & 80    & 20    & 0.95  & 0.4   & 0.7   & 2500  & 2500& 0.00  & 11.69 & 0.21  & 0\% \\
    80    & 80    & 80    & 80    & 80    & 80    & 0.925 & 0.4   & 0.7   & 6080  & 6080& 0.00  & 11.92 & 0.65  & 0\% \\
    80    & 80    & 80    & 80    & 80    & 80    & 0.95  & 0.4   & 0.7   & 4000  & 4000& 0.00  & 12.34 & 1.13  & 0\% \\
   \hline
\end{tabular}
}
\end{table}

\clearpage
\section{Dataset of Schools in Calabria}
\label{app:schools}

\scriptsize
\vspace{-0.5cm}

\begin{table}[htbp]
\centering
\caption{Overview of all the schools located in the Province of Catanzaro.}
\label{tab:cz}
\small %
\resizebox{\textwidth}{!}{
\begin{tabular}{cc c c || c cc c c}
\toprule
\textbf{School ID} & \textbf{Municipality} & \textbf{Type} & \textbf{Students} & & \textbf{School ID} & \textbf{Municipality} & \textbf{Type} & \textbf{Students} \\
\midrule
CZIC813004 & Serrastretta & CI & 490 & & CZIC86500R & Tiriolo & CI & 640 \\
CZIC81400X & Martirano & CI & 366 & & CZIC86700C & Catanzaro & CI & 1284 \\
CZIC81500Q & S. Mannelli & CI & 373 & & CZIC868008 & Lamezia Terme & CI & 1229 \\
CZIC818007 & Badolato & CI & 690 & & CZIC869004 & Soverato & CI & 793 \\
CZIC821003 & Davoli & CI & 820 & & CZIC87200X & Squillace & CI & 716 \\
CZIC82200V & Curinga & CI & 577 & & CZIC87300Q & Taverna & CI & 466 \\
CZIC82400E & Cropani & CI & 882 & & CZIC87400G & Lamezia Terme & CI & 1651 \\
CZIC82500A & Falerna & CI & 1136 & & CZIS001002 & Catanzaro & USI & 1473 \\
CZIC82900N & Maida & CI & 705 & & CZIS00200T & Girifalco & USI & 435 \\
CZIC83000T & Settingiano & CI & 665 & & CZIS00300N & Decollatura & USI & 356 \\
CZIC835001 & Sersale & CI & 796 & & CZIS007001 & Chiaravalle C. & USI & 505 \\
CZIC839008 & Borgia & CI & 791 & & CZIS01100L & Sersale & USI & 450 \\
CZIC84000C & Girifalco & CI & 574 & & CZIS01800B & Soverato & USI & 963 \\
CZIC842004 & Botricello & CI & 634 & & CZIS019007 & Lamezia Terme & USI & 1764 \\
CZIC84300X & Chiaravalle C. & CI & 1022 & & CZIS021007 & Catanzaro & USI & 1208 \\
CZIC84400Q & Lamezia Terme & CI & 1359 & & CZIS022004 & Catanzaro & USI & 1664 \\
CZIC84600B & Montepaone & CI & 646 & & CZIS02300V & Lamezia Terme & USI & 651 \\
CZIC848003 & Sellia Marina & CI & 779 & & CZIS02400P & Catanzaro & USI & 1089 \\
CZIC850003 & Lamezia Terme & CI & 1135 & & CZIS02600A & Lamezia Terme & USI & 1661 \\
CZIC85200P & Catanzaro & CI & 1608 & & CZPS02000R & Lamezia Terme & USI & 1031 \\
CZIC856002 & Catanzaro & CI & 1524 & & CZRH04000Q & Soverato & USI& 471 \\
CZIC85800N & Catanzaro & CI & 1619 & & CZTF010008 & Catanzaro & USI & 1383 \\
CZIC86000N & Catanzaro & CI & 901 & & CZTL06000D & Soverato & USI & 657 \\
CZIC864001 & Lamezia Terme & CI & 1568 & & CZVC01000A & Catanzaro & BSI & 1452 \\
\bottomrule
\end{tabular}}
\end{table}

\begin{table}[htbp]
\vspace{-0.5cm}
\centering
\caption{Overview of all the schools located in the Province of Vibo Valentia.}%
\label{tab:vv}
\small %
\resizebox{\textwidth}{!}{
\begin{tabular}{cc c c || c cc c c}
\toprule
\textbf{School ID} & \textbf{Municipality} & \textbf{Type} & \textbf{Students} & & \textbf{School ID} & \textbf{Municipality} & \textbf{Type} & \textbf{Students} \\
\midrule
VVIC803004 & Acquaro & CI & 932 & & VVIS00200C & Tropea & USI & 827 \\
VVIC81200V & San Costantino Calabro & CI & 1118 & & VVIS003008 & Serra San Bruno & USI & 641 \\
VVIC81300P & Rombiolo & CI & 1631 & & VVIS00700G & Vibo Valentia & USI& 704 \\
VVIC82000T & Sant'onofrio & CI & 650 & & VVIS011007 & Vibo Valentia & USI & 1249 \\
VVIC82200D & Tropea & CI & 1402 & & VVIS012003 & Vibo Valentia & USI& 1143 \\
VVIC824005 & Serra San Bruno & CI & 900 & & VVMM008008 & Filadelfia & II & 764 \\
VVIC82600R & Vibo Valentia & CI & 1318 & & VVPC04000D & Nicotera & II & 680 \\
VVIC831008 & Vibo Valentia & CI & 1512 & & VVPM01000T & Vibo Valentia & USI & 1362 \\
VVIC83300X & Pizzo & II & 1049 & & VVPS01000R & Vibo Valentia & USI & 1249 \\
VVIC83500G & Vallelonga & CI & 708 & & VVVC010001 & Vibo Valentia & BSI & 848 \\
\bottomrule
\end{tabular}}
\end{table}

\begin{table}[htbp]
\centering
\caption{Overview of all the schools located in the Province of Reggio di Calabria.}
\label{tab:rc}
\small %
\resizebox{\textwidth}{!}{
\begin{tabular}{cc c c || c cc c c}
\toprule
\textbf{School ID} & \textbf{Municipality} & \textbf{Type} & \textbf{Students} & & \textbf{School ID} & \textbf{Municipality} & \textbf{Type} & \textbf{Students} \\
\midrule
RCIC80200C & San Giorgio Morgeto & CI & 383 & & RCIC86600B & Siderno & CI & 900 \\
RCIC804004 & Reggio Di Calabria & CI & 914 & & RCIC867007 & Reggio Di Calabria & CI & 1339 \\
RCIC80500X & Reggio Di Calabria & CI & 1090 & & RCIC868003 & Reggio Di Calabria & CI & 801 \\
RCIC80600Q & Reggio Di Calabria & CI & 1201 & & RCIC86900V & Reggio Di Calabria & CI & 1648 \\
RCIC809007 & Reggio Di Calabria & CI & 1521 & & RCIC87100V & Reggio Di Calabria & CI & 1163 \\
RCIC812003 & Montebello Jonico & CI & 698 & & RCIC87200P & Reggio Di Calabria & CI & 1019 \\
RCIC81300V & M. di Gioiosa Ionica & CI & 806 & & RCIC87300E & Reggio Di Calabria & CI & 1224 \\
RCIC81400P & San Luca & CI & 489 & & RCIC87400A & Taurianova & CI & 1507 \\
RCIC81500E & Ardore & CI & 734 & & RCIS00100R & Melito Di Porto Salvo & USI & 474 \\
RCIC81600A & Gerace & CI & 288 & & RCIS00300C & Polistena & USI & 543 \\
RCIC817006 & Delianuova & CI & 410 & & RCIS00700Q & Bovalino & USI & 656 \\
RCIC81900T & S. Eufemia D'Aspr. & CI & 648 & & RCIS013003 & Gioia Tauro & USI& 1164 \\
RCIC82100T & Palmi & CI & 790 & & RCIS01400V & Rosarno & USI & 987 \\
RCIC825005 & Rosarno & CI & 853 & & RCIS01600E & Bova Marina & USI & 705 \\
RCIC826001 & Caulonia & CI & 1144 & & RCIS01700A & Bagnara Calabra & USI & 787 \\
RCIC82900C & Monasterace & CI & 826 & & RCIS019002 & Palmi & USI & 1099 \\
RCIC832008 & Oppido Mamertina & CI & 703 & & RCIS02200T & Oppido Mamertina & USI & 750 \\
RCIC83400X & Campo Calabro & CI & 923 & & RCIS03100L & Siderno & USI& 1219 \\
RCIC83700B & Plati' & CI & 684 & & RCIS03200C & Palmi & USI & 925 \\
RCIC839003 & Gioiosa Ionica & CI & 989 & & RCIS034004 & Reggio Di Calabria & USI & 1117 \\
RCIC841003 & Melito Di P. Salvo & CI & 1080 & & RCIS03600Q & Villa San Giovanni & USI & 695 \\
RCIC84200V & Reggio Di Calabria & CI & 801 & & RCIS03800B & Roccella Ionica & USI & 1075 \\
RCIC84300P & Bagnara Calabra & CI & 807 & & RCIS039007 & Cittanova & USI & 1216 \\
RCIC84400E & Bianco & CI & 808 & & RCIS041007 & Locri & USI & 1739 \\
RCIC84500A & Bovalino & CI & 1017 & & RCIS042003 & Locri & USI & 844 \\
RCIC846006 & Cinquefrondi & CI & 747 & & RCPC050008 & Reggio Di Calabria & USI & 1095 \\
RCIC847002 & Cittanova & CI & 891 & & RCPM02000L & Locri & USI & 911 \\
RCIC84800T & Laureana Di Borr. & CI & 1180 & & RCPM04000T & Reggio Di Calabria & USI & 1324 \\
RCIC85000T & Polistena & CI & 1249 & & RCPM05000C & Polistena & USI & 645 \\
RCIC85100N & Rizziconi & CI & 666 & & RCPS010001 & Reggio Di Calabria & USI & 1816 \\
RCIC85200D & Bova Marina & CI & 1037 & & RCPS030006 & Reggio Di Calabria & USI & 1414 \\
RCIC853009 & Locri & CI & 1277 & & RCRH100001 & Villa San Giovanni & USI& 796 \\
RCIC855001 & Villa San Giovanni & CI & 1101 & & RCTD120008 & Reggio Di Calabria & USI & 978 \\
RCIC85800C & Rosarno & CI & 1043 & & RCTF030008 & Polistena & USI & 1216 \\
RCIC859008 & Gioia Tauro & CI & 1796 & & RCTF05000D & Reggio Di Calabria & USI & 1598 \\
RCIC861008 & Palmi & CI & 1068 & & RCVC010005 & Reggio Di Calabria & BSI & 1164 \\
RCIC86500G & Siderno & CI & 927 & & & & & \\
\bottomrule
\end{tabular}}
\end{table}

\begin{table}[htbp]
\centering
\caption{Overview of all the schools located in the Province of Crotone.}%
\label{tab:kr}
\small %
\resizebox{\textwidth}{!}{
\begin{tabular}{cc c c || c cc c c}
\toprule
\textbf{School ID} & \textbf{Municipality} & \textbf{Type} & \textbf{Students} & & \textbf{School ID} & \textbf{Municipality} & \textbf{Type} & \textbf{Students} \\
\midrule
KRIC80300C & Crotone & CI & 1591 & & KRIC827001 & Mesoraca & CI & 600 \\
KRIC804008 & Rocca Di Neto & CI & 714 & & KRIC83100L & Isola Di Capo Rizzuto & CI & 1746 \\
KRIC80800G & Scandale & CI & 431 & & KRIC83200C & Petilia Policastro & CI & 887 \\
KRIC80900B & Strongoli & II & 727 & & KRIS00200R & Cotronei & USI& 393 \\
KRIC81000G & Crotone & CI & 1329 & & KRIS00400C & Ciro' Marina & USI & 460 \\
KRIC81100B & Crotone & CI & 1210 & & KRIS006004 & Cutro & USI & 189 \\
KRIC813003 & Crotone & CI & 1785 & & KRIS00900G & Crotone & USI & 857 \\
KRIC81500P & Cotronei & CI & 474 & & KRIS013007 & Crotone & USI & 1241 \\
KRIC81700A & Verzino & CI & 628 & & KRIS014003 & Crotone & USI & 1217 \\
KRIC820006 & Ciro' & II & 524 & & KRPC02000L & Crotone & USI & 660 \\
KRIC821002 & Caccuri & CI & 415 & & KRPM010006 & Crotone & USI & 898 \\
KRIC82400D & Ciro' Marina & CI & 1615 & & KRPS010005 & Crotone & USI & 1064 \\
KRIC825009 & Santa Severina & II & 749 & & KRPS02000Q & Petilia Policastro & USI & 547 \\
KRIC826005 & Cutro & CI & 890 & & KRRH050009 & Isola Di Capo Rizzuto & USI & 244 \\
\bottomrule
\end{tabular}}
\end{table}

\begin{table}[htbp]
\centering
\caption{Overview of all the schools located in the Province of Cosenza.}%
\label{tab:cs}
\small %
\resizebox{\textwidth}{!}{
\begin{tabular}{cc c c || c cc c c}
\toprule
\textbf{School ID} & \textbf{Municipality} & \textbf{Type} & \textbf{Students} & & \textbf{School ID} & \textbf{Municipality} & \textbf{Type} & \textbf{Students} \\
\midrule
CSIC80200T & Carolei & CI & 438 & & CSIC8AE00X & San Marco Argentano & CI & 659 \\
CSIC81100L & Corigliano-Rossano & CI & 1516 & & CSIC8AH00B & Corigliano-Rossano & CI & 1304 \\
CSIC81200C & Cosenza & CI & 1294 & & CSIC8AJ00L & Bisignano & CI & 743 \\
CSIC814004 & San Sosti & CI & 487 & & CSIC8AK00C & Rende & CI & 1905 \\
CSIC81500X & Fagnano Castello & CI & 599 & & CSIC8AL008 & Cosenza & CI & 1219 \\
CSIC81800B & Amantea & CI & 624 & & CSIC8AM004 & Corigliano-Rossano & CI & 896 \\
CSIC819007 & Belmonte Calabro & CI & 468 & & CSIC8AN00X & Corigliano-Rossano & CI & 1148 \\
CSIC822003 & Cassano All'ionio & CI & 1464 & & CSIC8AQ00B & Corigliano-Rossano & CI & 1034 \\
CSIC82300V & Villapiana & CI & 527 & & CSIC8AR007 & Crosia & CI & 978 \\
CSIC82400P & Francavilla Marittima & CI & 468 & & CSIC8AS00C & Belvedere Marittimo & CI & 788 \\
CSIC827006 & Morano Calabro & CI & 481 & & CSIC8AT008 & Tortora & CI & 484 \\
CSIC82900T & Mormanno & CI & 368 & & CSIC8AU004 & Praia A Mare & CI & 686 \\
CSIC83100T & Corigliano-Rossano & CI & 1556 & & CSIC8AV00X & San Giovanni In Fiore & CI & 462 \\
CSIC836001 & Diamante & CI & 554 & & CSIC8AX00G & San Giovanni In Fiore & CI & 767 \\
CSIC83700R & Santa Maria Del Cedro & CI & 809 & & CSIC8AY00B & Scalea & CI & 1004 \\
CSIC84100C & San Lucido & CI & 604 & & CSIS001006 & Corigliano-Rossano & USI & 866 \\
CSIC842008 & Terranova Da Sibari & CI & 762 & & CSIS014008 & Amantea & USI & 743 \\
CSIC84600G & Cropalati & CI & 250 & & CSIS01600X & Bisignano & USI & 486 \\
CSIC848007 & Longobucco & II & 287 & & CSIS01700Q & Cosenza & USI& 1180 \\
CSIC849003 & Mandatoriccio & CI & 471 & & CSIS01800G & Acri & USI & 1248 \\
CSIC850007 & Rocca Imperiale & CI & 410 & & CSIS022007 & Cassano All'ionio & USI & 522 \\
CSIC851003 & Mangone & CI & 982 & & CSIS023003 & Diamante & USI & 498 \\
CSIC85200V & Amendolara & CI & 432 & & CSIS02700A & Roggiano Gravina & USI & 522 \\
CSIC85400E & Casali Del Manco & CI & 852 & & CSIS028006 & Cetraro & USI & 966 \\
CSIC85500A & Spezzano Della Sila & CI & 722 & & CSIS04600Q & Corigliano-Rossano & USI & 800 \\
CSIC857002 & San Pietro In Guarano & CI & 742 & & CSIS049007 & Castrolibero & USI & 866 \\
CSIC85800T & Lungro & II & 624 & & CSIS051007 & Cosenza & USI & 1264 \\
CSIC86100N & Guardia Piemontese & CI & 458 & & CSIS06300D & Trebisacce & USI & 834 \\
CSIC864005 & Scigliano & CI & 185 & & CSIS064009 & Corigliano-Rossano & USI & 1076 \\
CSIC865001 & Amantea & CI & 877 & & CSIS06700R & San Marco Argentano & USI & 689 \\
CSIC86700L & Torano Castello & CI & 447 & & CSIS06800L & Cariati & USI & 771 \\
CSIC87000C & Mendicino & CI & 599 & & CSIS07100C & Corigliano-Rossano & USI & 1022 \\
CSIC871008 & Paola & CI & 116 & & CSIS072008 & Paola & USI & 1370 \\
CSIC872004 & Cetraro & CI & 789 & & CSIS073004 & Cosenza & USI & 839 \\
CSIC87300X & Fuscaldo & CI & 607 & & CSIS07700B & San Giovanni In Fiore & USI & 424 \\
CSIC87400Q & Rogliano & CI & 665 & & CSIS078007 & San Giovanni In Fiore & USI & 543 \\
CSIC87500G & Roggiano Gravina & CI & 960 & & CSIS079003 & Castrovillari & USI & 105 \\
CSIC87600B & Castrolibero & CI & 699 & & CSIS081003 & Cosenza & USI & 1265 \\
CSIC877007 & Cerisano & CI & 606 & & CSIS08200V & Scalea & USI & 802 \\
CSIC878003 & Spezzano Albanese & CI & 628 & & CSIS08300P & Cosenza & USI & 1545 \\
CSIC87900V & Rende & CI & 1126 & & CSIS08400E & Castrovillari & USI& 636 \\
CSIC88300E & Acri & CI & 752 & & CSIS086006 & Corigliano-Rossano & USI & 1206 \\
CSIC88700T & Montalto Uffugo & CI & 1045 & & CSIS087002 & Castrovillari & USI & 829 \\
CSIC88800N & Montalto Uffugo & CI & 1744 & & CSPC010007 & Cosenza & USI & 957 \\
CSIC89000N & Rende & CI & 1177 & & CSPC060008 & San Demetrio Corone & II & 654 \\
CSIC892009 & Cariati & CI & 772 & & CSPC190001 & Rende & USI & 653 \\
CSIC89600L & Cosenza & CI & 103 & & CSPM070003 & Belvedere Marittimo & USI & 523 \\
CSIC89700C & Cosenza & CI & 1003 & & CSPS03000G & Cosenza & USI & 1158 \\
CSIC898008 & Cosenza & CI & 1473 & & CSPS18000D & Rende & USI & 1474 \\
CSIC899004 & Acri & CI & 662 & & CSPS310001 & Trebisacce & USI & 652 \\
CSIC8A000R & Trebisacce & CI & 870 & & CSTF01000C & Cosenza & USI & 1264 \\
CSIC8A200C & Luzzi & CI & 608 & & CSVC01000E & Cosenza & BSI & 378 \\
CSIC8A3008 & Castrovillari & CI & 1618 & & & & & \\
\bottomrule
\end{tabular}}
\end{table}

\normalsize
\newpage

\section{Classical Optimization Results for the Calabria Case Study}
\label{app:calabria_results}

\begin{table}[htbp]
\centering
\caption{Computational results obtained with the classical optimization model for the 64 policy configurations of the Calabria case study. For each configuration, the table reports the objective coefficients ($C_1$, $C_2$, $C_3$, $C_5$, $C_6$), the autonomy threshold ($\gamma$), the optimal objective value ($z^{C}$), and the solution time. The maximum execution times were fixed to $T^{IC}_{\max}=1200$~s and $T^{IS}_{\max}=2400$~s.} \label{tab:calabria_classical_full} 
\small %
\resizebox{\textwidth}{!}{
\begin{tabular}{cccccccc||cccccccc}
\toprule
$C_1$ & $C_2$ & $C_3$ & $C_5$ & $C_6$ & $\gamma$ & $z^C$ & Time (s) & $C_1$ & $C_2$ & $C_3$ & $C_5$ & $C_6$ & $\gamma$ & $z^C$ & Time (s) \\ 
\midrule
20 & 20 & 20 & 20 & 20 & 0.925 & 940  & 0.04 & 80 & 20 & 20 & 20 & 20 & 0.925 & 1960 & 0.20 \\
20 & 20 & 20 & 20 & 20 & 0.950 & 560  & 0.16 & 80 & 20 & 20 & 20 & 20 & 0.950 & 1160 & 0.11 \\
20 & 20 & 20 & 20 & 80 & 0.925 & 1260 & 0.12 & 80 & 20 & 20 & 20 & 80 & 0.925 & 2280 & 0.13 \\
20 & 20 & 20 & 20 & 80 & 0.950 & 640  & 0.04 & 80 & 20 & 20 & 20 & 80 & 0.950 & 1240 & 0.05 \\
20 & 20 & 20 & 80 & 20 & 0.925 & 1000 & 0.04 & 80 & 20 & 20 & 80 & 20 & 0.925 & 2200 & 0.07 \\
20 & 20 & 20 & 80 & 20 & 0.950 & 580  & 0.15 & 80 & 20 & 20 & 80 & 20 & 0.950 & 1400 & 0.18 \\
20 & 20 & 20 & 80 & 80 & 0.925 & 1480 & 0.04 & 80 & 20 & 20 & 80 & 80 & 0.925 & 2520 & 0.05 \\
20 & 20 & 20 & 80 & 80 & 0.950 & 780  & 0.04 & 80 & 20 & 20 & 80 & 80 & 0.950 & 1480 & 0.04 \\
20 & 20 & 80 & 20 & 20 & 0.925 & 1240 & 0.04 & 80 & 20 & 80 & 20 & 20 & 0.925 & 2260 & 0.06 \\
20 & 20 & 80 & 20 & 20 & 0.950 & 560  & 0.04 & 80 & 20 & 80 & 20 & 20 & 0.950 & 1160 & 0.09 \\
20 & 20 & 80 & 20 & 80 & 0.925 & 1600 & 0.04 & 80 & 20 & 80 & 20 & 80 & 0.925 & 2620 & 0.06 \\
20 & 20 & 80 & 20 & 80 & 0.950 & 800  & 0.06 & 80 & 20 & 80 & 20 & 80 & 0.950 & 1400 & 0.05 \\
20 & 20 & 80 & 80 & 20 & 0.925 & 1420 & 0.05 & 80 & 20 & 80 & 80 & 20 & 0.925 & 2500 & 0.03 \\
20 & 20 & 80 & 80 & 20 & 0.950 & 620  & 0.08 & 80 & 20 & 80 & 80 & 20 & 0.950 & 1400 & 0.08 \\
20 & 20 & 80 & 80 & 80 & 0.925 & 1780 & 0.06 & 80 & 20 & 80 & 80 & 80 & 0.925 & 2860 & 0.04 \\
20 & 20 & 80 & 80 & 80 & 0.950 & 980  & 0.05 & 80 & 20 & 80 & 80 & 80 & 0.950 & 1640 & 0.05 \\
20 & 80 & 20 & 20 & 20 & 0.925 & 1300 & 0.07 & 80 & 80 & 20 & 20 & 20 & 0.925 & 2320 & 0.05 \\
20 & 80 & 20 & 20 & 20 & 0.950 & 680  & 0.04 & 80 & 80 & 20 & 20 & 20 & 0.950 & 1440 & 0.04 \\
20 & 80 & 20 & 20 & 80 & 0.925 & 2200 & 0.06 & 80 & 80 & 20 & 20 & 80 & 0.925 & 3220 & 0.03 \\
20 & 80 & 20 & 20 & 80 & 0.950 & 1400 & 0.04 & 80 & 80 & 20 & 20 & 80 & 0.950 & 2000 & 0.05 \\
20 & 80 & 20 & 80 & 20 & 0.925 & 1340 & 0.05 & 80 & 80 & 20 & 80 & 20 & 0.925 & 2560 & 0.06 \\
20 & 80 & 20 & 80 & 20 & 0.950 & 680  & 0.06 & 80 & 80 & 20 & 80 & 20 & 0.950 & 1520 & 0.04 \\
20 & 80 & 20 & 80 & 80 & 0.925 & 2260 & 0.04 & 80 & 80 & 20 & 80 & 80 & 0.925 & 3460 & 0.04 \\
20 & 80 & 20 & 80 & 80 & 0.950 & 1420 & 0.04 & 80 & 80 & 20 & 80 & 80 & 0.950 & 2240 & 0.06 \\
20 & 80 & 80 & 20 & 20 & 0.925 & 1720 & 0.07 & 80 & 80 & 80 & 20 & 20 & 0.925 & 2740 & 0.08 \\
20 & 80 & 80 & 20 & 20 & 0.950 & 920  & 0.04 & 80 & 80 & 80 & 20 & 20 & 0.950 & 1520 & 0.05 \\
20 & 80 & 80 & 20 & 80 & 0.925 & 2500 & 0.06 & 80 & 80 & 80 & 20 & 80 & 0.925 & 3520 & 0.06 \\
20 & 80 & 80 & 20 & 80 & 0.950 & 1400 & 0.06 & 80 & 80 & 80 & 20 & 80 & 0.950 & 2000 & 0.06 \\
20 & 80 & 80 & 80 & 20 & 0.925 & 1860 & 0.08 & 80 & 80 & 80 & 80 & 20 & 0.925 & 2980 & 0.07 \\
20 & 80 & 80 & 80 & 20 & 0.950 & 940  & 0.05 & 80 & 80 & 80 & 80 & 20 & 0.950 & 1760 & 0.08 \\
20 & 80 & 80 & 80 & 80 & 0.925 & 2680 & 0.06 & 80 & 80 & 80 & 80 & 80 & 0.925 & 3760 & 0.05 \\
20 & 80 & 80 & 80 & 80 & 0.950 & 1460 & 0.04 & 80 & 80 & 80 & 80 & 80 & 0.950 & 2240 & 0.05 \\
   \hline
\end{tabular}
}
\end{table}

\begin{table}[ht]
\centering
\caption{Summary of the main structural indicators obtained over the 64 policy configurations for the Calabria case study. For each province, the table reports the average value together with the minimum and maximum values observed across all experiments.}
\label{tab:solution_analysis}
\small
\resizebox{\textwidth}{!}{
\begin{tabular}{lccccc} 
\toprule
\textbf{Indicator} & \textbf{CZ} & \textbf{KR} & \textbf{RC} & \textbf{CS} & \textbf{VV}\\
\midrule

Cross-municipality
& 3.70 (2--5)
& 2.34 (0--4)
& 2.81 (2--3)
& 5.58 (2--9)
& 0.00 (0--0) \\

Compatible tracks
& 2.00 (0--3)
& 1.08 (0--4)
& 1.19 (0--2)
& 3.98 (0--7)
& 0.00 (0--0) \\

Incompatible tracks
& 2.36 (0--4)
& 1.27 (0--4)
& 2.44 (0--3)
& 3.19 (0--5)
& 0.00 (0--0) \\

Educational hubs
& 4.36 (3--6)
& 2.34 (0--4)
& 3.63 (2--4)
& 7.17 (4--10)
& 0.00 (0--0) \\

Average criticality
& 1.31 (0.67--2.00)
& 1.37 (0.00--2.75)
& 1.07 (1.00--1.50)
& 1.38 (0.75--2.00)
& 0.00 (0.00--0.00) \\

\bottomrule
\end{tabular}}
\end{table}

\FloatBarrier

\section{Quantum Optimization Results for the Calabria Case Study}
\label{quantum_table_case}

\begin{table}[ht] 
\centering
\caption{Computational results obtained with the hybrid quantum optimization framework for the Calabria case study. For each policy configuration, the table reports the objective-function coefficients ($C_1$, $C_2$, $C_3$, $C_5$, and $C_6$), the autonomy threshold ($\gamma$), the optimal objective value obtained with the classical optimization model ($z^{C}$), the average objective value over five independent quantum executions ($\overline{z}^{Q}$), the standard deviation of the quantum objective values ($\sigma_{z}^{Q}$), the average quantum solution time ($\overline{t}^{Q}$), the corresponding standard deviation ($\sigma_{t}^{Q}$), and the optimality gap with respect to the classical optimum.}
\label{tab:calabria_quantum_full}
\smallskip
\resizebox{0.96\textwidth}{!}{ 
\begin{tabular}{cccccc|c|cccc|c || ccccccc|c|cccc|c}
\toprule
\multicolumn{6}{c|}{\textbf{Parameters}} & \textbf{Classical} & \multicolumn{4}{c|}{\textbf{Quantum}} & \multirow{2}{*}{\textbf{Gap}} & &
\multicolumn{6}{c|}{\textbf{Parameters}} & \textbf{Classical} & \multicolumn{4}{c|}{\textbf{Quantum}} & \multirow{2}{*}{\textbf{Gap}} \\
\cmidrule(lr){1-6} \cmidrule(lr){7-7} \cmidrule(lr){8-11} 
\cmidrule(lr){14-19} \cmidrule(lr){20-20} \cmidrule(lr){21-24}

$C_1$ & $C_2$ & $C_3$ & $C_5$ & $C_6$ & $\gamma$ & $z^{C}$ & $\overline{z}^{Q}$ & $\sigma_{z}^{Q}$ & $\overline{t}^{Q}$ & $\sigma_{t}^{Q}$ & & &
$C_1$ & $C_2$ & $C_3$ & $C_5$ & $C_6$ & $\gamma$ & $z^{C}$ & $\overline{z}^{Q}$ & $\sigma_{z}^{Q}$ & $\overline{t}^{Q}$ & $\sigma_{t}^{Q}$ & \\
\midrule
20 & 20 & 20 & 20 & 20 & 0.95  & 560  & 560  & 0.00 & 10.47 & 0.44 & 0\% & & 80 & 20 & 20 & 20 & 20 & 0.925 & 1960 & 1960 & 0.00 & 12.85 & 2.98 & 0\% \\
20 & 20 & 20 & 20 & 80 & 0.925 & 1260 & 1260 & 0.00 & 10.45 & 0.27 & 0\% & & 80 & 20 & 20 & 20 & 20 & 0.95  & 1160 & 1160 & 0.00 & 10.68 & 0.65 & 0\% \\
20 & 20 & 20 & 20 & 80 & 0.95  & 640  & 640  & 0.00 & 10.14 & 0.13 & 0\% & & 80 & 20 & 20 & 20 & 80 & 0.925 & 2280 & 2280 & 0.00 & 10.90 & 0.51 & 0\% \\
20 & 20 & 20 & 80 & 20 & 0.925 & 1000 & 1000 & 0.00 & 10.04 & 0.04 & 0\% & & 80 & 20 & 20 & 20 & 80 & 0.95  & 1240 & 1240 & 0.00 & 11.24 & 1.24 & 0\% \\
20 & 20 & 20 & 80 & 20 & 0.95  & 580  & 580  & 0.00 & 10.19 & 0.20 & 0\% & & 80 & 20 & 20 & 80 & 20 & 0.925 & 2200 & 2200 & 0.00 & 10.97 & 0.53 & 0\% \\
20 & 20 & 20 & 80 & 80 & 0.925 & 1480 & 1480 & 0.00 & 10.21 & 0.28 & 0\% & & 80 & 20 & 20 & 80 & 20 & 0.95  & 1400 & 1400 & 0.00 & 10.71 & 0.44 & 0\% \\
20 & 20 & 20 & 80 & 80 & 0.95  & 780  & 780  & 0.00 & 10.02 & 0.15 & 0\% & & 80 & 20 & 20 & 80 & 80 & 0.925 & 2520 & 2520 & 0.00 & 10.42 & 0.33 & 0\% \\
20 & 20 & 80 & 20 & 20 & 0.925 & 1240 & 1240 & 0.00 & 11.44 & 2.19 & 0\% & & 80 & 20 & 80 & 20 & 20 & 0.925 & 2260 & 2260 & 0.00 & 10.62 & 0.40 & 0\% \\
20 & 20 & 80 & 20 & 20 & 0.95  & 560  & 560  & 0.00 & 10.12 & 0.09 & 0\% & & 80 & 20 & 80 & 20 & 20 & 0.95  & 1160 & 1160 & 0.00 & 11.18 & 1.11 & 0\% \\
20 & 20 & 80 & 20 & 80 & 0.925 & 1600 & 1600 & 0.00 & 10.23 & 0.35 & 0\% & & 80 & 20 & 80 & 20 & 80 & 0.925 & 2620 & 2620 & 0.00 & 10.97 & 0.54 & 0\% \\
20 & 20 & 80 & 20 & 80 & 0.95  & 800  & 800  & 0.00 & 10.44 & 0.51 & 0\% & & 80 & 20 & 80 & 20 & 80 & 0.95  & 1400 & 1400 & 0.00 & 11.29 & 0.70 & 0\% \\
20 & 20 & 80 & 80 & 20 & 0.925 & 1420 & 1420 & 0.00 & 10.35 & 0.46 & 0\% & & 80 & 20 & 80 & 80 & 20 & 0.925 & 2500 & 2500 & 0.00 & 10.88 & 0.80 & 0\% \\
20 & 20 & 80 & 80 & 20 & 0.95  & 620  & 620  & 0.00 & 10.46 & 0.60 & 0\% & & 80 & 20 & 80 & 80 & 20 & 0.95  & 1400 & 1400 & 0.00 & 11.45 & 1.41 & 0\% \\
20 & 20 & 80 & 80 & 80 & 0.925 & 1780 & 1780 & 0.00 & 10.79 & 1.05 & 0\% & & 80 & 20 & 80 & 80 & 80 & 0.925 & 2860 & 2860 & 0.00 & 11.13 & 0.93 & 0\% \\
20 & 20 & 80 & 80 & 80 & 0.95  & 980  & 980  & 0.00 & 10.67 & 1.23 & 0\% & & 80 & 20 & 80 & 80 & 80 & 0.95  & 1640 & 1640 & 0.00 & 10.67 & 0.73 & 0\% \\
20 & 80 & 20 & 20 & 20 & 0.925 & 1300 & 1300 & 0.00 & 11.12 & 1.81 & 0\% & & 80 & 80 & 20 & 20 & 20 & 0.925 & 2320 & 2320 & 0.00 & 11.81 & 1.33 & 0\% \\
20 & 80 & 20 & 20 & 20 & 0.95  & 680  & 680  & 0.00 & 10.45 & 0.57 & 0\% & & 80 & 80 & 20 & 20 & 20 & 0.95  & 1440 & 1440 & 0.00 & 10.63 & 0.18 & 0\% \\
20 & 80 & 20 & 20 & 80 & 0.925 & 2200 & 2200 & 0.00 & 12.15 & 2.38 & 0\% & & 80 & 80 & 20 & 20 & 80 & 0.925 & 3220 & 3220 & 0.00 & 11.06 & 0.43 & 0\% \\
20 & 80 & 20 & 20 & 80 & 0.95  & 1400 & 1400 & 0.00 & 10.17 & 0.05 & 0\% & & 80 & 80 & 20 & 20 & 80 & 0.95  & 2000 & 2000 & 0.00 & 10.75 & 0.64 & 0\% \\
20 & 80 & 20 & 80 & 20 & 0.925 & 1340 & 1340 & 0.00 & 10.19 & 0.05 & 0\% & & 80 & 80 & 20 & 80 & 20 & 0.925 & 2560 & 2560 & 0.00 & 10.37 & 0.21 & 0\% \\
20 & 80 & 20 & 80 & 20 & 0.95  & 680  & 680  & 0.00 & 10.18 & 0.33 & 0\% & & 80 & 80 & 20 & 80 & 20 & 0.95  & 1520 & 1520 & 0.00 & 11.27 & 0.66 & 0\% \\
20 & 80 & 20 & 80 & 80 & 0.925 & 2260 & 2260 & 0.00 & 10.62 & 0.74 & 0\% & & 80 & 80 & 20 & 80 & 80 & 0.925 & 3460 & 3460 & 0.00 & 11.52 & 0.95 & 0\% \\
20 & 80 & 20 & 80 & 80 & 0.95  & 1420 & 1420 & 0.00 & 10.30 & 0.11 & 0\% & & 80 & 80 & 20 & 80 & 80 & 0.95  & 2240 & 2240 & 0.00 & 10.63 & 0.50 & 0\% \\
20 & 80 & 80 & 20 & 20 & 0.925 & 1720 & 1720 & 0.00 & 10.32 & 0.33 & 0\% & & 80 & 80 & 80 & 20 & 20 & 0.925 & 2740 & 2740 & 0.00 & 11.16 & 1.01 & 0\% \\
20 & 80 & 80 & 20 & 20 & 0.95  & 920  & 920  & 0.00 & 10.50 & 0.49 & 0\% & & 80 & 80 & 80 & 20 & 20 & 0.95  & 1520 & 1520 & 0.00 & 11.23 & 0.74 & 0\% \\
20 & 80 & 80 & 20 & 80 & 0.925 & 2500 & 2500 & 0.00 & 11.08 & 2.00 & 0\% & & 80 & 80 & 80 & 20 & 80 & 0.925 & 3520 & 3520 & 0.00 & 11.25 & 1.13 & 0\% \\
20 & 80 & 80 & 20 & 80 & 0.95  & 1400 & 1400 & 0.00 & 11.53 & 1.98 & 0\% & & 80 & 80 & 80 & 20 & 80 & 0.95  & 2000 & 2000 & 0.00 & 10.93 & 0.64 & 0\% \\
20 & 80 & 80 & 80 & 20 & 0.925 & 1860 & 1860 & 0.00 & 11.14 & 1.90 & 0\% & & 80 & 80 & 80 & 80 & 20 & 0.925 & 2980 & 2980 & 0.00 & 10.72 & 0.60 & 0\% \\
20 & 80 & 80 & 80 & 20 & 0.95  & 940  & 940  & 0.00 & 10.20 & 0.18 & 0\% & & 80 & 80 & 80 & 80 & 20 & 0.95  & 1760 & 1760 & 0.00 & 10.81 & 0.85 & 0\% \\
20 & 80 & 80 & 80 & 80 & 0.925 & 2680 & 2680 & 0.00 & 10.26 & 0.32 & 0\% & & 80 & 80 & 80 & 80 & 80 & 0.925 & 3760 & 3760 & 0.00 & 10.26 & 0.23 & 0\% \\
20 & 80 & 80 & 80 & 80 & 0.95  & 1460 & 1460 & 0.00 & 10.33 & 0.45 & 0\% & & 80 & 80 & 80 & 80 & 80 & 0.95  & 2240 & 2240 & 0.00 & 10.45 & 0.12 & 0\% \\
\bottomrule
\end{tabular}
}
\end{table}

\end{document}